\documentclass[12pt]{article}
\usepackage{graphicx}
\catcode`\@=11
\def\numberbysection{\@addtoreset{equation}{section}
\def\theequation{\thesection.\arabic{equation}}}

\numberbysection
\newcommand{\abs}[1]{\vert#1\vert}
\newcommand{\act}[1]{\underline{#1}}
\newcommand{\ap}{_{\rm ap}}
\newcommand{\aps}{_{\rm ap}^\star}
\newcommand{\be}{\[}
\newcommand{\bea}{\begin{eqnarray*}}
\newcommand{\beq}{\begin{equation}}
\newcommand{\beqa}{\begin{eqnarray}}
\newcommand{\bin}[2]{{#1\choose#2}}
\newcommand{\bl}{\circ}
\newcommand{\ee}{\]}
\newcommand{\eea}{\end{eqnarray*}}
\newcommand{\eeq}{\end{equation}}
\newcommand{\eeqa}{\end{eqnarray}}
\newcommand{\con}{{\rm conn}}
\newcommand{\cum}[1]{\langle\!\langle#1\rangle\!\rangle}
\renewcommand{\d}{{\rm d}}
\newcommand{\diag}{{\rm diag}}
\newcommand{\ds}[1]{{\displaystyle #1}}
\newcommand{\e}{{\rm e}}

\newcommand{\eq}{{\rm eq}}
\newcommand{\frad}[2]{\displaystyle{\displaystyle#1\over\displaystyle#2}}
\renewcommand{\i}{{\rm i}}
\newcommand{\lam}{\lambda}
\newcommand{\lra}{\Longleftrightarrow}
\newcommand{\elam}{\e^\lam}
\newcommand{\emam}{\e^{-\lam}}
\newcommand{\mean}[1]{\langle#1\rangle}
\newcommand{\bigmean}[1]{\left\langle#1\right\rangle}
\newcommand{\no}{\bullet}
\newcommand{\p}{p}
\newcommand{\s}{\sigma}
\newcommand{\sqr}{\sqrt\ds}
\newcommand{\st}{{^\star}}
\renewcommand{\t}{\tau}
\newcommand{\var}{\mathop{\rm Var}}
\newcommand\vd{{\vphantom{\ds{M}}}}
\newcommand\vm{{\vphantom{{M}}}}
\newcommand{\A}{{\cal A}}
\newcommand{\B}{{\cal B}}
\newcommand{\C}{{\cal C}}
\newcommand{\D}{{\cal D}}
\newcommand{\G}{\Gamma}
\renewcommand{\H}{{\cal H}}
\newcommand{\N}{{\cal N}}
\renewcommand{\P}{{\cal P}}
\renewcommand{\S}{\Sigma}
\newcommand{\Sp}{{\cal S}}
\newcommand{\T}{{\cal T}}
\newcommand{\W}{{\cal W}}

\begin{document}
\centerline{\Large\bf Jamming, freezing and metastability}
\vspace{.3cm}
\centerline{\Large\bf in one-dimensional spin systems}
\vspace{1.6cm}
\centerline{\large
G.~De Smedt$^{a,}$\footnote{desmedt@spht.saclay.cea.fr},
C.~Godr\`eche$^{b,}$\footnote{godreche@spec.saclay.cea.fr},
and J.M.~Luck$^{a,}$\footnote{luck@spht.saclay.cea.fr}}
\vspace{1cm}
\centerline{$^a$Service de Physique Th\'eorique\footnote{URA 2306 of CNRS},
CEA Saclay, 91191 Gif-sur-Yvette cedex, France}
\centerline{$^b$Service de Physique de l'\'Etat Condens\'e,
CEA Saclay, 91191 Gif-sur-Yvette cedex, France}
\vspace{1cm}

\begin{abstract}
We consider in parallel
three one-dimensional spin models with kinetic constraints:
the paramagnetic constrained Ising chain,
the ferromagnetic Ising chain with constrained Glauber dynamics,
and the same chain with constrained Kawasaki dynamics.
At zero temperature the dynamics of these models is fully irreversible,
leading to an exponentially large number of blocked states.
Using a mapping of these spin systems onto sequential adsorption models of,
respectively, monomers, dimers, and hollow trimers,
we present exact results on the statistics of blocked states.
We determine the distribution of their energy or magnetization,
and in particular the large-deviation function
describing its exponentially small tails.
The spin and energy correlation functions are also determined.
The comparison with an approach based on a priori statistics
reveals systematic discrepancies with the Edwards hypothesis,
concerning in particular the fall-off of correlations.
\end{abstract}

\vfill
\noindent To appear in European Physical Journal B

\noindent P.A.C.S.: 05.70.Ln, 64.60.My, 68.43.Fg, 68.43.Mn, 75.40.Gb.

\newpage
\setcounter{footnote}{0}
\section{Introduction}

In a great variety of systems,
such as structural glasses, spin glasses, and granular materials,
the dynamics at low temperature or high density
is so slow that the system falls out of equilibrium~\cite{ange}.
For long, glassy dynamics has been described
as a slow motion in a complex energy (or free energy) landscape~\cite{gold},
with many valleys separated by barriers.
Several approaches have been proposed
in order to make this heuristic picture more precise.
Valleys thus appear under various names (and with various definitions)
in different contexts:
metastable states, TAP states~\cite{tap}, pure states~\cite{ktw},
inherent structures~\cite{sw}, quasi-states~\cite{fv}.
From a dynamical viewpoint these concepts are not equivalent~\cite{bir}.
Metastability is indeed unambiguously defined for mean-field models only,
where metastable states have infinite lifetimes,
as barrier heights diverge with the system size.
For finite-dimensional systems
with short-range interactions,
barrier heights and valley lifetimes are always finite at finite temperature,
so that metastability becomes a matter of time scales~\cite{gs}.

Once these valleys are appropriately defined,
one can estimate their number at fixed energy (or free energy) density $E$.
This number generically grows exponentially with the system size, as
\beq
\N(N)\sim\exp(N S\ap(E)),
\label{entropy}
\eeq
where $S\ap(E)$ is the configurational entropy or complexity.
The subscript `ap' (a priori) refers to the fact that, when counting
valleys, each of them appears with the same weight.
In this {\it a priori} ensemble, valleys are {\it combinatorially} equivalent.

In contrast,
a key question concerns the {\it dynamical} weight of each individual valley.
Do all the valleys play a similar role in the dynamics?
This question arises for instance
when a system is instantaneously quenched into the glassy phase,
starting from a disordered configuration.
Assuming this initial configuration is chosen at random,
does the system sample all the possible valleys with a given final energy $E$
with equal statistical weights, i.e., with a uniform or flat measure,
or, to the contrary,
does the size of the attraction basin of each valley really matter?
The same question is also relevant in another situation
commonly referred to as tapping~\cite{tam1,tapping,tam2}.
Under tapping a granular material continuously jumps
from a blocked configuration to a nearby one.
Does the non-equilibrium steady state thus obtained
admit a statistical description
in terms of a flat ensemble of blocked configurations?

The answer to this question for the first situation
(relaxational or aging dynamics)
is positive at least in some mean-field models,
where valleys are known to be explored with a flat measure~\cite{fv}.
The concept of ergodicity, and the resulting thermodynamical construction,
therefore hold, as in equilibrium situations,
up to the replacement of configurations by valleys.
The configurational temperature $T\ap$, defined by
\beq
\frac{1}{T\ap}=\frac{\d S\ap}{\d E},
\label{tapri}
\eeq
has a thermodynamical meaning.
It also coincides with the effective temperature
involved in the generalized fluctuation-dissipation formula
in the appropriate temporal regime.

Besides the mean-field geometry,
another physical situation where metastable states
are unambiguously defined is the zero-temperature limit,
where no barrier can be crossed at all.
Valleys are just blocked configurations under the chosen dynamics.
For instance, for an Ising model with single-spin dynamics,
a valley is a configuration where each spin is aligned with its local field,
provided the latter is non-zero.

In the context of granular materials,
Edwards~\cite{edwards} proposed to describe the slow compaction dynamics
by means of a flat ensemble average
over all the blocked configurations of the grains with prescribed density.
Extending the range of application of this idea,
the so-called Edwards hypothesis consists in assuming
that all the valleys with a given energy density are equivalent.
This hypothesis has two consequences.
First, the value of an observable can be obtained by a flat average
over the a priori ensemble, or Edwards ensemble, of all those valleys.
Second, the temperature $T\ap$ of~(\ref{tapri}),
also known as the Edwards temperature,
has the usual thermodynamical meaning of a temperature.

The present paper is devoted to the analytical study
of the zero-temperature dynamics of simple one-dimensional
systems with kinetic constraints.
Kinetically constrained models have been the subject
of numerous investigations~\cite{barc,scic,scictau,acic,acictau,ldean,
prbr,priv,lint,kra,cri,buga,bfs,lefe}.
Here we specialize to Ising chains without frustration nor
quenched disorder, namely
paramagnetic chains (CIC)~\cite{scic,scictau,acic,acictau} (section~2)
and ferromagnetic chains~\cite{ldean,prbr,priv,lint,kra} (sections~3 and~4).
The common feature of all these models is the irreversible nature
of their zero-temperature dynamics:
each spin flips at most once in the whole history of the system.

The central goal of this paper is to obtain exact results on the statistics
of the blocked configurations reached by these systems,
pursuing the efforts made by the authors of some recent
works~\cite{ldean,prbr,priv,lint,kra,cri,buga,bfs,lefe}.
Thus doing we are able to critically revisit the questions raised above.
In the derivation of these results
we will take advantage of the fact that the zero-temperature dynamics of these
models can be rephrased in terms of random sequential adsorption (RSA)
or cooperative sequential adsorption (CSA)~\cite{rsa},
for which analytical techniques are available in one dimension.
A blocked configuration thus appears as a jammed state of the corresponding RSA
or CSA model.
The Edwards hypothesis is,
in this specific context, akin to the question raised long
ago~\cite{widom} whether RSA configurations are equilibrated or not.
Our results confirm that the answer is negative.
A more extensive discussion will be given in section~5.

\section{Constrained Ising chain}
\subsection{Definition of the model}

Constrained Ising chains (CIC)~\cite{scic,scictau,acic,acictau}
are among the simplest examples of kinetically constrained models~\cite{barc}.
Although they have trivial equilibrium properties,
the presence of kinetic constraints leads to slow dynamics at low temperature,
and to metastability at zero temperature.

Consider a paramagnetic Ising chain, made of independent spins $\s_n=\pm1$,
submitted to a positive unit magnetic field.
This model has a Hamiltonian
\beq
\H=-\sum_n\s_n,
\label{ham}
\eeq
with a unique ordered ground state, where all the spins are up ($\s_n=+1$).

Kinetic constraints are introduced as follows.
Consider single-spin-flip dynamics with rates
\be
W(\s_n\to-\s_n)=\min(1,\e^{-2\beta\s_n})\,W_0(\s_{n-1},\s_{n+1}).
\ee
The first factor in the right side is the Metropolis acceptance rate,
ensuring detailed balance with respect to the Hamiltonian $\H$
at temperature $T=1/\beta$.
The second factor imposes a kinetic constraint:
the flipping rate of a spin $\s_n$ depends on its environment,
i.e., on the value of its neighbors $\s_{n-1}$ and $\s_{n+1}$.
Let us make the choice~\cite{cri}
\beq
W_0(\s_{n-1},\s_{n+1})=a\t_{n-1}+(1-a)\t_{n+1},
\label{w0}
\eeq
with the notation
\be
\t_n=\frac{1-\s_n}{2}.
\ee

The parameter $0\le a\le1$ allows to interpolate between known limiting cases.
For $a=1/2$, the constraint factor is $(\t_{n-1}+\t_{n+1})/2$,
i.e., half the number of neighboring down spins.
The symmetrically constrained chain~(SCIC)~\cite{scic,scictau}
is thus obtained.
For $a=0$, the constraint factor is $\t_{n+1}$:
the spin $\s_n$ can only flip if its right neighbor is down.
The right asymmetrically constrained chain~(ACIC)~\cite{acic,acictau}
is thus recovered.
Similarly, the left ACIC is obtained for $a=1$.

At zero temperature, the dynamics of the CIC simplifies drastically.
An initially up spin ($\s_n=+1$) remains up forever,
while a down spin ($\s_n=-1$) can flip at most once, according to
stochastic rules depending on its two neighbors:
\beq
\left\{\matrix{
---\to-+-&&(\hbox{rate}\ 1),\hfill\cr
--+\to-++&&(\hbox{rate}\ a),\hfill\cr
+--\to++-&&(\hbox{rate}\ 1-a).\hfill\cr
}\right.
\label{spinrule}
\eeq

For a finite chain of $N$ spins,
the dynamics stops after a finite jamming time $T_N$,
which depends both on the initial configuration
and on the history of the chain.
The jamming time will be shown to grow as $T_N\approx\ln N$,
up to finite fluctuations given by extreme-value statistics.
The system is thus left after a finite time $T_N$ in a jammed or blocked state.
This state, which is an attractor for the dynamics,
is a spin configuration where each down spin is isolated,
i.e., surrounded by two up spins.
The blocked configuration thus obtained depends on the parameter~$a$,
on the initial configuration of the chain, and on its whole stochastic history.

The problem may be equivalently described as
an irreversible process of particle adsorption.
Consider indeed down spins as representing empty sites ($\circ$),
and up spins as representing occupied sites ($\bullet$):
\be
\left\{\matrix{
\s=-1\lra\t=1\lra\bl,\hfill\cr
\s=+1\lra\t=0\lra\no.\hfill
}\right.
\ee
The zero-temperature dynamics of the CIC thus maps onto a problem
of particle adsorption, where individual particles (monomers)
are irreversibly deposited according to:
\be
\left\{\matrix{
\bl\bl\bl\to\bl\no\bl&&(\hbox{rate}\ 1),\hfill\cr
\bl\bl\no\to\bl\no\no&&(\hbox{rate}\ a),\hfill\cr
\no\bl\bl\to\no\no\bl&&(\hbox{rate}\ 1-a).\hfill\cr
}\right.
\ee
The deposition rate at site $n$ depends on the occupation state
of both neighboring sites.
We are thus facing a cooperative sequential adsorption (CSA) model~\cite{rsa}.
The limit (jamming) coverage of this model
is related to the mean magnetization per spin $M(\infty)$
in the blocked configurations by
\be
\P_\infty(\no)=1-\mean{\tau}_\infty=\frac{1+M(\infty)}{2}.
\ee
In the following, we will use the language of spins, magnetization,
spin correlations, and so on,
leaving the picture of particle deposition for illustrative purposes only.

\subsection{A priori statistics}

We have shown that the attractors of the zero-temperature dynamics of the CIC,
for any value of the parameter $a$, are the spin configurations
where each down spin is isolated, i.e., surrounded by two up spins.

A natural statistical description of these attractors
is provided by the a priori ensemble,
or Edwards ensemble, as explained in the Introduction,
where all the blocked spin configurations are taken with equal weights.

For a finite chain of $N$ spins,
consider the restricted ensemble of blocked configurations
for which exactly $n$ spins are down.
Their magnetization $M$ is such that $NM=N-2n$,
with $0\le n\le N/2$, hence $0\le M\le1$.
The number of such configurations reads
\beq
\N(N,n)=\bin{N-n+1}{n}.
\label{bino}
\eeq
Indeed this is the number of ways of inserting $n$ down spins
in the $N-n+1$ spaces made available by the presence
of $N-n$ up spins, with at most one down spin per space.
This number grows exponentially, according to~(\ref{entropy}),
where the a priori entropy reads~\cite{cri,bfs,ldean}
\beq
S\ap(M)=-M\ln(2M)+\frac{1+M}{2}\ln(1+M)-\frac{1-M}{2}\ln(1-M).
\label{sm}
\eeq

One can also consider the full (or unrestricted) ensemble
of all the blocked configurations, irrespective of their magnetization.
The number $\N(N)$ of such configurations can be determined as follows.
For a chain of $N\ge3$ spins, a configuration either ends with $(+-)$
(there are $\N(N-2)$ such configurations)
or with $(+)$ (there are $\N(N-1)$ such configurations).
We thus obtain the recursion relation $\N(N)=\N(N-1)+\N(N-2)$,
with $\N(1)=2$, $\N(2)=3$, hence
\be
\N(N)=F_{N+2},
\ee
where $F_N$ are the Fibonacci numbers.
This expression is also equal to the sum of (\ref{bino}) for $n$
ranging from $0$ to $N/2$.
It grows as $\N(N)\sim\exp(NS\aps)$, with
\beq
S\aps=\ln\Phi=0.481212,
\label{s}
\eeq
where $\Phi=(1+\sqrt{5})/2$ is the golden mean.
The result~(\ref{s}) is the maximum value of the function
$S\ap(M)$~(\ref{sm}).
This maximum is reached for
\beq
M\st=\frac{1}{\sqrt{5}}=0.447214,
\label{m}
\eeq
which is therefore the typical a priori magnetization
of a blocked configuration.

The distribution of the number $n$ of down spins
in the a priori ensemble is given by
\be
P_n=\frac{\N(N,n)}{\N(N)}.
\ee
For a large sample ($N\gg1$), the probability density of the magnetization $M$
is therefore given by an exponential estimate of the form
\beq
f(M)\sim\exp(-N\,\S\ap(M)),
\label{sig}
\eeq
with
\beq
\S\ap(M)=S\aps-S\ap(M).
\label{ssig}
\eeq
The result~(\ref{sig}) has the form of large-deviation estimates
in probability theory,
which hold e.g.~for the arithmetic mean of $N$ independent random variables.
The large-deviation function (or entropy function) $\S\ap(M)$
will be plotted in Figure~\ref{f4}.
It vanishes quadratically near $M=M\st$ as
\be
\S\ap(M)\approx c\left(M-M\st\right)^2,\qquad c=\frac{5\sqrt{5}}{8}.
\ee
The bulk of the a priori distribution of $M$
is therefore asymptotically a narrow Gaussian around $M\st$,
with a scaled variance given by $N\var M\approx1/(2c)=4\sqrt{5}/25=0.357771$.

The a priori entropy can alternatively be evaluated
by the transfer-matrix method~\cite{tm}.
For a finite chain of $N$ spins,
we introduce the characteristic function of the magnetization
\be
Z_N(\beta)=\sum_\C\e^{\beta NM(\C)},
\ee
where the sum runs over all the blocked configurations $\C$.
Note that $NM(\C)$ is the opposite of the total energy
of the configuration, according to the Hamiltonian~(\ref{ham}),
so that $Z_N(\beta)$ coincides with the usual
partition function of the model, at a fictitious inverse temperature $\beta$.

The partition functions $Z_N^\pm$ of a finite chain of $N$ spins,
labeled by the prescribed value $\s_N=\pm1$ of the last spin,
obey the recursion
\be
\pmatrix{Z_{N+1}^+\cr Z_{N+1}^-}=\T\pmatrix{Z_N^+\cr Z_N^-},
\ee
where the $2\times2$ transfer matrix
\be
\T=\pmatrix{\e^\beta&\e^\beta\cr\e^{-\beta}&0}
\ee
has eigenvalues
\be
\lam_\pm(\beta)=\frac{\e^\beta\pm\sqr{4+\e^{2\beta}}}{2}.
\ee

The entropy $S\ap(M)$ is then given by a Legendre transform.
We have indeed
\be
Z_N(\beta)\sim\int\e^{N(S\ap(M)+\beta M)}\,\d M\sim\e^{N\ln\lam_+(\beta)}.
\ee
Evaluating the integral by the steepest-descent method
yields the `thermodynamical' relationships
\beq
\ln\lam_+(\beta)-S\ap(M)=\beta M,
\qquad M=\frac{\d\ln\lam_+}{\d\beta},
\qquad\beta=-\frac{\d S\ap}{\d M},
\label{thertrans}
\eeq
which yield in the present case
\be
M=\frac{\e^\beta}{\sqr{4+\e^{2\beta}}},\qquad\e^\beta=\frac{2M}{\sqr{1-M^2}},
\ee
and allow to recover~(\ref{sm}).

The spin correlation function $C_n=\mean{\s_0\s_n}$ can also be evaluated
in the a priori ensemble at fixed magnetization
by the transfer-matrix method~\cite{tm}.
We have, for $n\ge0$ in the bulk of an infinitely long chain,
\be
C_n=\frad{\langle L_+\vert\Sp\T^n\Sp\vert R_+\rangle}{\lam_+^n}
=\left(\langle L_+\vert\Sp\vert R_+\rangle\right)^2
+\langle L_+\vert\Sp\vert R_-\rangle\langle L_-\vert\Sp\vert R_+\rangle
\left(\frac{\lam_-}{\lam_+}\right)^n.
\ee
In this expression, $\Sp=\diag(+1,-1)$ is the spin operator, while
\be
\langle L_\pm\vert=\frac{1}{\lam_\pm^2+1}\pmatrix{\lam_\pm&\e^\beta},\qquad
\vert R_\pm\rangle=\pmatrix{\e^\beta\lam_\pm\cr1}
\ee
are the left and right eigenvectors of $\T$
associated with the eigenvalues $\lam_\pm$.
We have consistently $M=\langle L_+\vert\Sp\vert R_+\rangle$.
After some algebra we obtain the following expression,
involving only the magnetization~$M$~\cite{ldean}:
\beq
C_n^\con=C_n-M^2=(1-M^2)\left(-\frac{1-M}{1+M}\right)^n.
\label{cr}
\eeq
The connected correlation function thus exhibits an exponential decay,
modulated by an oscillating sign.

The full ensemble of blocked configurations
is obtained by setting $\beta=0$ in the above results,
which indeed corresponds to taking a flat average
over all blocked configurations.
This prescription amounts to replacing the magnetization $M$
by its typical value $M\st$~(\ref{m}).
We thus obtain in particular
\beq
C_n^\con=C_n-\frac{1}{5}=\frac{4}{5}\left(-\frac{1}{\Phi^2}\right)^n.
\label{cunr}
\eeq

We end up by mentioning that the blocked spin configurations
considered so far are the degenerate ground states
of the antiferromagnetic Ising chain in a constant magnetic field
$h=2J>0$~\cite{abc},
whose Hamiltonian reads
\be
\H=J\sum_n\s_n\s_{n+1}-2J\sum_n\s_n.
\ee
As a consequence, the above expressions are exact results
for the latter model at equilibrium at zero temperature.
This is one of the simplest models with
a non-zero entropy at zero temperature, given by~(\ref{s}).

\subsection{Dynamics of cluster densities and magnetization}

We now turn to the exact analysis of the zero-temperature dynamics of the CIC,
starting with the mean cluster densities and magnetization.

We consider an uncorrelated magnetized initial state, given by
\beq
\left\{\matrix{
\s_n(0)=-1,\hfill&\t_n(0)=1\hfill&(\bl)\hfill
&\hbox{with prob.}\;\p,\hfill\cr
\s_n(0)=+1,\hfill&\t_n(0)=0\hfill&(\no)\hfill
&\hbox{with prob.}\;1-\p,\hfill
}\right.
\label{eps}
\eeq
so that the mean initial magnetization reads $M(0)=1-2\p$.

For $\p\le1/2$, the initial state~(\ref{eps}) is the equilibrium state
of the Ising chain with Hamiltonian~(\ref{ham}) at inverse temperature
\beq
\beta_0=\frac{1}{2}\,\ln\frac{1-\p}{\p}.
\label{beta0}
\eeq
In particular, a random (unmagnetized) initial configuration, i.e., $p=1/2$,
corresponds to infinite temperature, i.e., $\beta_0=0$.

It is a common feature of one-dimensional RSA and similar problems~\cite{rsa}
that the densities of certain patterns, including active clusters,
obey closed rate equations.
Consider clusters of exactly~$\ell\ge1$ consecutive down spins.
Their density per unit length at time $t$ reads
\beq
p_\ell(t)
=\mean{(1-\t_0)\t_1\dots\t_\ell(1-\t_{\ell+1})}_t
=\P_t(\no\underbrace{\bl\cdots\bl}_{\ell}\no),
\label{pt}
\eeq
and the mean magnetization of the chain is given by
\beq
M(t)=1-2\sum_{\ell\ge1}\ell\,p_\ell(t).
\label{mp}
\eeq
Because zero-temperature dynamics is fully irreversible,
the densities $p_\ell(t)$ obey rate equations, which can be derived as follows.
Clusters of length $\ell=1$ are inactive.
Consider a cluster of length $\ell\ge2$,
renumbering its sites as $n=1,\dots,\ell$.
The spin $\s_n$ can flip from down to up, at a rate given by~(\ref{spinrule}),
thus generating one or two smaller clusters of the following length
\beq
\left\{\matrix{
n=1\hfill&(\hbox{rate}\ 1-a),\hfill&
\quad\hbox{one cluster:}\hfill&\ell_1=\ell-1,\hfill\cr
2\le n\le\ell-1\hfill&(\hbox{rate}\ 1),\hfill&
\quad\hbox{two clusters:}\hfill&
\ell_1=n-1,\;\ell_2=\ell-n,\hfill\cr
n=\ell\hfill&(\hbox{rate}\ a),\hfill&
\quad\hbox{one cluster:}\hfill&\ell_1=\ell-1.\hfill
}\right.
\label{cut}
\eeq

Gathering the contributions of all these events,
we obtain the rate equations
\beq
\frac{\d p_\ell(t)}{\d t}
=-(\ell-1)p_\ell(t)+p_{\ell+1}(t)+2\sum_{k\ge\ell+2}p_k(t)
\label{prat}
\eeq
for $\ell\ge1$, irrespective of the value of the asymmetry parameter $a$.
The initial state~(\ref{eps}) yields $p_\ell(0)=(1-\p)^2\p^\ell$.

A simple way of solving the rate equations~(\ref{prat})
consists in making the Ansatz
\beq
p_\ell(t)=a(t)\,z(t)^\ell
\label{pan}
\eeq
for $\ell\ge1$.
We obtain successively $\d z(t)/\d t=-z(t)$, with $z(0)=\p$, hence
\beq
z(t)=\p\e^{-t},
\label{z}
\eeq
and $\d a(t)/\d t=a(t)(1+z(t)^2)/(1-z(t))$, with $a(0)=(1-\p)^2$, hence
\be
a(t)=\e^t(1-\p\e^{-t})^2\exp(\p(\e^{-t}-1)),
\ee
so that finally
\beq
p_\ell(t)=(1-\p\e^{-t})^2\exp(\p(\e^{-t}-1))\,\p^\ell\e^{-(\ell-1)t}.
\label{psol}
\eeq
As expected, only inactive clusters of length $\ell=1$
survive in the final states, and their density reads
\be
p_1(\infty)=\p\e^{-\p}.
\ee
Equation~(\ref{mp}) yields
\beq
M(t)=1-2\p\exp(\p(\e^{-t}-1)),
\label{mrest}
\eeq
and especially
\beq
M(\infty)=1-2\p\e^{-\p}.
\label{mresinf}
\eeq

\begin{figure}[htb]
\begin{center}
\includegraphics[angle=90,width=.7\linewidth]{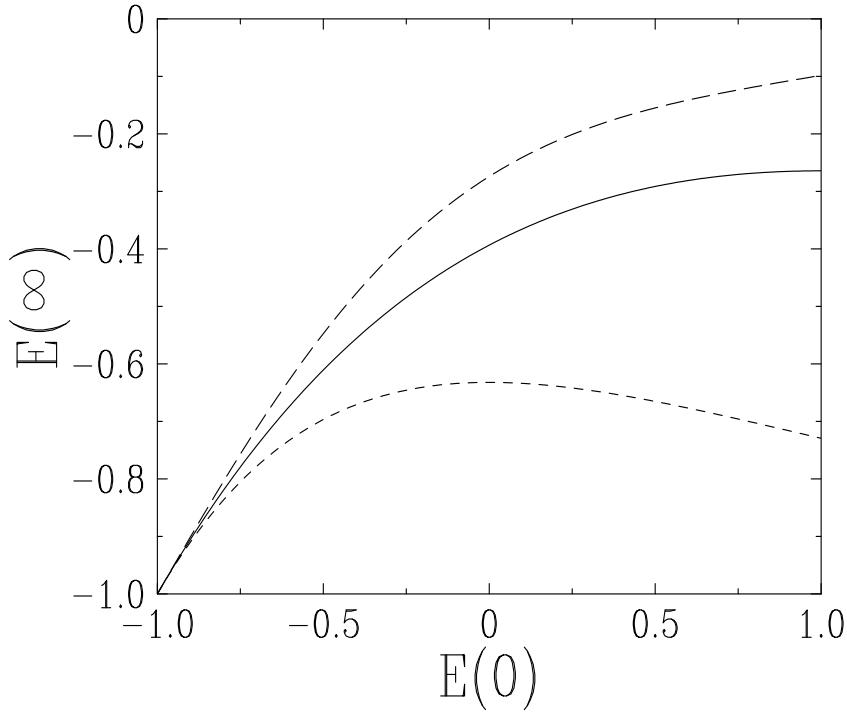}
\caption{\small
Plot of the final energy $E(\infty)$ against the initial energy $E(0)$.
Full line: CIC~(\ref{mresinf}).
Dashed line: ferromagnetic chain with constrained
Glauber dynamics~(\ref{efinf}).
Long-dashed line: ferromagnetic chain with
constrained Kawasaki dynamics~(\ref{ekinf}).}
\label{f1}
\end{center}
\end{figure}

The mean final magnetization of the blocked states reached by the dynamics
thus depends on the parameter $\p$ characterizing the initial state.
This non-trivial dependence demonstrates that the dynamics is not ergodic.
For an initial state close to the ground-state ($\p\to0$, i.e., $M(0)\to1$),
the behavior $M(\infty)\approx M(0)+2\p^2$
is easily explained in terms of clusters of two down spins:
the density of these clusters scales as $\p^2$,
and only one of the two spins will flip.
For a random (unmagnetized) initial configuration ($\p=1/2$, i.e., $M(0)=0$),
we have
\beq
M(\infty)_{\p=1/2}=1-\e^{-1/2}=0.393469.
\label{minfran}
\eeq
As this number is the final magnetization of a typical initial state,
it is natural to compare it to the prediction~(\ref{m})
of the a priori ensemble.
This comparison will be presented in Table~1.
Finally, for the ordered initial state where all spins
are down ($\p=1$, i.e., $M(0)=-1$),
we have the smallest possible value of the final magnetization:
\beq
M(\infty)_{\p=1}=1-2\e^{-1}=0.264241.
\label{minfmin}
\eeq
Figure~\ref{f1} shows a plot of the final energy $E(\infty)=-M(\infty)$
against the initial one, $E(0)=-M(0)$,
for the present model, as well as for the ferromagnetic chain with
constrained Glauber and Kawasaki dynamics~(see sections~3 and~4).

\begin{figure}[htb]
\begin{center}
\includegraphics[angle=90,width=.7\linewidth]{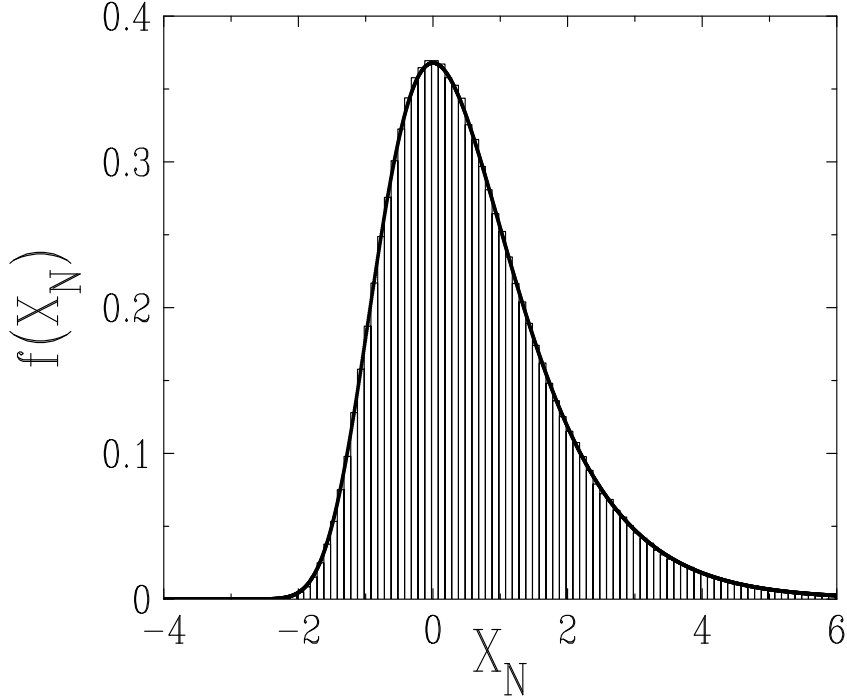}
\caption{\small
Distribution of the jamming time $T_N$ of CIC samples of $N=1000$ spins,
against $X_N$ defined in~(\ref{tx}).
Histogram: numerical data $(p=1/2)$.
Thick line: limit Gumbel law~(\ref{gum}).}
\label{f2}
\end{center}
\end{figure}

To close up, we present an analysis of the distribution
of the jamming time $T_N$ of a large but finite system of $N$ spins,
a question which does not seem to have been considered in previous works
on RSA.
Equation~(\ref{psol}) shows that the late stages of the dynamics
are governed by an exponentially small density of surviving clusters
made of two down spins, $p_2(t)\approx\alpha\e^{-t}$, with $\alpha=p^2\e^{-p}$.
The dynamics can therefore be effectively described by
a collection of $\alpha N$ such clusters,
each cluster decaying exponentially with unit rate, when a down spin flips.
The jamming time $T_N$ is the largest of the decay times of those clusters.
For a large sample, it is therefore distributed according to extreme-value
statistics~\cite{gumbel}.
Setting
\beq
T_N=\ln(\alpha N)+X_N,
\label{tx}
\eeq
we find that the fluctuation $X_N$ remains of order unity,
and that it is asymptotically distributed according to the Gumbel law
\beq
f(X)=\exp(-X-\e^{-X}).
\label{gum}
\eeq
We have checked this prediction by a numerical simulation.
Figure~\ref{f2} shows a histogram of the observed jamming time $T_N$
for $10^6$ samples of $N=1000$ spins,
starting with a random initial configuration.
The bin size is $\Delta T=1/10$.
An excellent agreement is found with the limit law~(\ref{gum}),
with $p=1/2$, hence $\alpha=\e^{-1/2}/4=0.151633$.

\subsection{Spin correlations}

The time-dependent spin correlation function reads
\beq
C_n(t)=\mean{\s_0\s_n}_t=\mean{(1-2\t_0)(1-2\t_n)}_t=1-4c_1(t)+4d_{1,n-1,1}(t),
\label{cord}
\eeq
where we have introduced the one-cluster function $c_n(t)$
and the two-cluster function $d_{m,k,n}(t)$, defined as
\beqa
&&c_n(t)=\mean{\t_1\dots\t_n}_t
=\P_t(\underbrace{\bl\cdots\bl}_n),\label{cn}\\
&&d_{m,k,n}(t)=\mean{\t_1\dots\t_m\,\t_{m+k+1}\dots\t_{m+k+n}}_t
=\P_t(\underbrace{\bl\cdots\bl}_{m}\;\underbrace{\cdots}_{k}
\;\underbrace{\bl\cdots\bl}_{n}).\label{dn}
\eeqa

The one-cluster function is the probability that the sites $1,\dots,n$
belong to a cluster of at least $n$ consecutive down spins.
It is therefore directly related to $p_n(t)$ defined in~(\ref{pt}) by
\beq
p_n(t)=c_n(t)-2c_{n+1}(t)+c_{n+2}(t).
\label{pc}
\eeq
Equation~(\ref{pan}) then yields
\beq
c_n(t)=A(t)\,z(t)^n,
\label{cna}
\eeq
with~(\ref{z}) and
\beq
A(t)=\frac{a(t)}{(1-\p\e^{-t})^2}=\e^t\exp(\p(\e^{-t}-1)).
\label{bt}
\eeq
This result can be alternatively recovered
by deriving rate equations for the one-cluster function itself.
As a consequence of~(\ref{w0}),
each variable $\t_2,\dots,\t_{n-1}$ entering the definition~(\ref{cn})
can flip from 1 to 0 with rate $c_n(t)$ per unit time,
while the rate for the first variable $\t_1$ is $ac_n(t)+(1-a)c_{n+1}(t)$,
and the rate for the last variable $\t_n$ is $ac_{n+1}(t)+(1-a)c_n(t)$.
Gathering all the contributions, we obtain the equation
\beq
\frac{\d c_n(t)}{\d t}=-(n-1)c_n(t)-c_{n+1}(t),
\label{crat}
\eeq
whose solution coincides with~(\ref{cna}),~(\ref{bt}).

A similar analysis yields the following rate equations
for the two-cluster function:
\beqa
&&{\hskip -.5cm}\frac{\d d_{m,k,n}(t)}{\d t}=-(m+n-2)d_{m,k,n}(t)\label{drat}\\
&&{\hskip 1.75cm}-a(d_{m+1,k,n}(t)+d_{m,k-1,n+1}(t))
-(1-a)(d_{m+1,k-1,n}(t)+d_{m,k,n+1}(t))\nonumber
\eeqa
for $m,k,n\ge1$,
with initial conditions $d_{m,k,n}(0)=\p^{m+n}$,
and boundary values $d_{m,0,n}(t)=c_{m+n}(t)$.
The rate equations~(\ref{drat}) are solved by the Ansatz
\beq
d_{m,k,n}(t)=B_k(t)\,z(t)^{m+n},
\label{dna}
\eeq
provided the amplitudes $B_k(t)$ obey
\beq
\frac{\d B_k(t)}{\d t}=(2-z(t))B_k(t)-z(t)B_{k-1}(t)
\label{ddrat}
\eeq
for $k\ge1$, irrespective of the value of $a$,
with the initial condition $B_k(0)=1$,
and the boundary value $B_0(t)=A(t)$~(\ref{bt}).

In order to solve~(\ref{ddrat}), we introduce the generating series
\beq
\B(x,t)=\sum_{k\ge1}B_k(t)x^k,
\label{genera}
\eeq
which obeys the differential equation
\be
\frac{\d\B(x,t)}{\d t}=(2-(x+1)z(t))\B(x,t)-xz(t)A(t),
\ee
considering $x$ as a parameter, with initial condition $\B(x,0)=x/(1-x)$.
This equation can be solved by `varying the constant':
\beq
\B(x,t)=\e^{2t}\exp(\p(\e^{-t}-1))
\left[\left(\frac{1}{\p x}+\frac{1}{1-x}\right)\exp(\p x(\e^{-t}-1))
-\frac{1}{\p x}-\e^{-t}\right].
\label{dres}
\eeq
Inserting~(\ref{cna}) and~(\ref{dna}) into~(\ref{cord}), we obtain
\be
C_n(t)=1-4\p\exp(\p(\e^{-t}-1))+4\p^2\e^{-2t}B_{n-1}(t).
\ee
Finally, expanding~(\ref{dres}) and using~(\ref{mrest}),
we obtain the expression of the connected spin correlation function:
\bea
&&C^\con_n(t)=C_n(t)-M(t)^2\nonumber\\
&&{\hskip 1.425cm}=4\p\exp(\p(\e^{-t}-1))
\left((1-\p)\frac{(\p(\e^{-t}-1))^n}{n!}
-\p\sum_{m\ge n+1}\frac{(\p(\e^{-t}-1))^m}{m!}\right).
\eea
As a consequence, in the blocked states, the correlation function reads
\beq
C^\con_n(\infty)=C_n(\infty)-M(\infty)^2=4\p\e^{-\p}
\left((1-\p)\frac{(-\p)^n}{n!}-\p\sum_{m\ge n+1}\frac{(-\p)^m}{m!}\right).
\label{cresinf}
\eeq

\begin{figure}[htb]
\begin{center}
\includegraphics[angle=90,width=.7\linewidth]{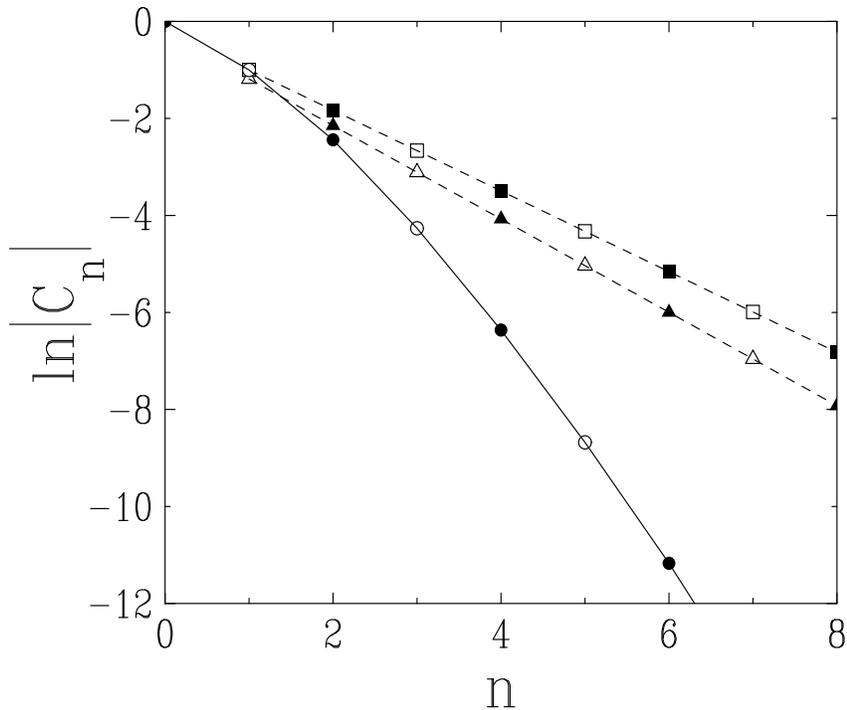}
\caption{\small
Connected spin correlation function in the final states of the CIC.
Full (open) symbols show positive (negative) correlations.
Circles and full line: logarithm of $(-1)^n$ times
the connected correlation $C^\con_n(\infty)$~(\ref{cresinf})
for $\p=1/2$, against $n$.
Squares and dashed line: prediction~(\ref{cr}),~(\ref{minfran})
of the restricted a priori ensemble.
Triangles and dashed line: prediction~(\ref{cunr})
of the full a priori~ensemble.}
\label{f3}
\end{center}
\end{figure}

The first term in the above expressions is the leading one,
implying that the connected correlation function
has a factorial asymptotic decay of the form $(\p(1-\e^{-t}))^n/n!$,
modulated by an oscillating sign, for any value of $\p$ and any time $t$.
This super-exponential fall-off is a characteristic feature
of irreversible processes such as RSA~\cite{rsa}.
This behavior is entirely missed
by the a priori approach~(\ref{cr}),~(\ref{cunr}),
where correlations fall off exponentially,
as they generically do in equilibrium systems.
Figure~\ref{f3} shows a logarithmic plot of $(-1)^n\,C^\con_n(\infty)$
against $n$, for a random initial configuration $(\p=1/2)$,
together with both predictions of the a priori approach,
i.e., the full ensemble~(\ref{cunr}), and the restricted ensemble~(\ref{cr})
where the exact magnetization~(\ref{minfran}) is imposed.
Both predictions appear as straight lines on the plot.
The exact value of $C^\con_1(\infty)$ is correctly reproduced
in the restricted a priori ensemble.

\subsection{Distribution of final magnetization and dynamical entropy}

We finally determine the full distribution of the number of spin flips
and of the final magnetization, for a given finite sample.
This problem was tackled long ago by a somewhat
similar approach~\cite{core} in the case of dimer deposition,
without consideration of the dynamical entropy, though.

\subsubsection*{Single active cluster}

We consider first the case of a single active cluster of size $\ell\ge2$,
with free boundary conditions, with all spins being initially down.
In the language of deposition, this corresponds to an initially empty cluster.
We are interested in the distribution of the number $\nu_\ell$
of spin flips (i.e., deposited particles) during the history of this cluster,
until it reaches a blocked configuration.
The final magnetization $M_\ell$ of the cluster is such that
\be
\ell M_\ell=\sum_{n=1}^\ell\s_n(\infty)=2\nu_\ell-\ell.
\ee

Because of the irreversible character of the dynamics,
every spin flip replaces the cluster where it takes place
by one or two smaller clusters, according to~(\ref{cut}).
In the generic case, two clusters of lengths $\ell_1$ and $\ell_2$
are generated, and their subsequent histories are independent.
We have therefore
\beq
\nu_\ell=1+\nu_{\ell_1}+\nu_{\ell_2},
\label{nucut}
\eeq
where $\nu_{\ell_1}$ and $\nu_{\ell_2}$ are independent random variables,
whose distribution is to be determined, while $\ell_1=n-1$ and $\ell_2=\ell-n$,
with the breaking point $n$ being uniform in the range $2\le n\le\ell-1$.
For $n=1$ or $n=\ell$, only one cluster is generated,
and~(\ref{nucut}) is changed accordingly.
Finally, we set $\nu_0=0$, which necessarily holds,
and $\nu_1=0$, which contains the gist of the kinetic constraint in the CIC
model.
Let
\beq
\phi_\ell(\lam)=\mean{\e^{\lam\nu_\ell}}
\label{carac}
\eeq
be the characteristic function of the distribution of $\nu_\ell$.
Equation~(\ref{nucut}) implies
\be
(\ell-1)\phi_\ell(\lam)
=\elam\sum_{k=1}^{\ell-1}\phi_k(\lam)\phi_{\ell-k-1}(\lam)
\ee
for $\ell\ge2$, with $\phi_0(\lam)=\phi_1(\lam)=1$.
These quadratic recursion relations can be solved by introducing
the generating series
\beq
\Phi(x,\lam)=\sum_{\ell\ge0}\phi_\ell(\lam)x^\ell,
\label{phidef}
\eeq
which obeys
\beq
x\frac{\d\Phi(x,\lam)}{\d x}=(\Phi(x,\lam)-1)(1+x\elam\Phi(x,\lam)),
\label{ricc}
\eeq
with $\Phi(x,\lam)=1+x+\cdots$ as $x\to0$.

The quadratic differential equation~(\ref{ricc})
has an obvious solution $\Phi(x,\lam)=1$.
Setting $\Phi(x,\lam)=1+1/u(x,\lam)$, we obtain a linear equation
\be
x\frac{\d u(x,\lam)}{\d x}+(1+x\elam)u(x,\lam)=-x\elam,
\ee
which can be solved by `varying the constant'.
This yields
\beq
\Phi(x,\lam)=\frac{\exp(x\elam)+\elam-1}{(1-x\elam)\exp(x\elam)+\elam-1}.
\label{phi}
\eeq
This expression formally contains
the distribution of the number $\nu_\ell$ of spin flips.

First, by expanding~(\ref{phi}) around $\lam=0$,
we obtain generating series for the successive moments of $\nu_\ell$.
The first of these series,
\be
\sum_{\ell\ge0}\mean{\nu_\ell}x^\ell=\frac{x(1-\e^{-x})}{(1-x)^2},
\ee
can be inverted explicitly, yielding
\beq
\mean{\nu_\ell}=(1-\e^{-1})\ell-\e^{-1}
+(\ell+1)\sum_{m\ge\ell+2}\frac{(-1)^m}{m!}.
\label{numean}
\eeq
The mean number of spin flips therefore grows linearly
with the cluster size $\ell$, with a coefficient $F_1=1-\e^{-1}=0.632120$,
in agreement with~(\ref{minfmin}).
The constant~${}-\e^{-1}$ can be viewed as the contribution
of the free ends of the cluster,
while the last term falls off factorially, with an oscillating sign,
as $(-1)^\ell/(\ell+1)!$, just like
the connected spin correlation~(\ref{cresinf}) for $\p=1$.

Then, by inverting the generating series~(\ref{phidef}),
we obtain the exponential estimate
\beq
\phi_\ell(\lam)\sim\e^{\ell F(\lam)},
\label{caract}
\eeq
where $x_c(\lam)=\exp(-F(\lam))$ is the zero of the denominator of~(\ref{phi}).
As a consequence, all the cumulants of $\nu_\ell$
grow linearly with the size $\ell$ of the cluster, as
\be
\cum{\nu_\ell^k}\approx F_k\ell,
\ee
with
\be
F(\lam)=\sum_{k\ge1}\frac{F_k\lam^k}{k!},\qquad
F_k=\left(\frac{\d^k F}{\d\lam^k}\right)_{\lam=0}.
\ee
We recover the above result $\mean{\nu_\ell}\approx F_1\ell$,
whereas $\var\nu_\ell\approx F_2\ell$, with $F_2=3\e^{-2}-\e^{-1}=0.038126$.
The bulk of the distribution of $\nu_\ell$ is therefore
a Gaussian of the form
\beq
P(\nu_\ell)\approx(2\pi F_2\ell)^{-1/2}
\,\exp\!\left(-\frac{(\nu_\ell-F_1\ell)^2}{2F_2\ell}\right).
\label{abg}
\eeq
In order to investigate the tails of the distribution of $\nu_\ell$
for $\ell$ large, we set
\be
\xi=\frac{\nu_\ell}{\ell}=\frac{1+M_\ell}{2}.
\ee
An inverse Laplace transform of (\ref{carac}), using
(\ref{caract}), yields
\be
P(\nu_\ell)\sim\int\frac{\d\lam}{2\pi\i}\,\e^{\ell(F(\lam)-\lam\xi)}.
\ee
Evaluating this integral by the saddle-point method,
we obtain an exponential estimate similar to~(\ref{sig}):
\beq
P(\nu_\ell)\sim\exp(-\ell\,\S(\xi)),
\label{abl}
\eeq
where the functions $F(\lam)$ and $\S(\xi)$ are
related to each other by a Legendre transform:
\be
F(\lam)+\S(\xi)=\lam\xi,
\qquad\lam=\frac{\d\S}{\d\xi},\qquad\xi=\frac{\d F}{\d\lam}.
\ee
The function $\S(\xi)$ is the large-deviation function
(or entropy function) of the quantity $\nu_\ell$.
This is a positive, convex function of $\xi$,
which vanishes quadratically around $\mean{\xi}=F_1$, as
\be
\S(\xi)\approx\frac{(\xi-F_1)^2}{2F_2},
\ee
in agreement with the Gaussian law~(\ref{abg}).

Coming back to the language of the magnetization $M$,
the above functions have the following parametric form,
in terms of $z=x_c(\lam)\elam$:
\beq
\matrix{
F=\ln(1-(1-z)\e^z)-\ln z,\hfill&{\hskip 5.5pt}\lam=\ln(1-(1-z)\e^z),\hfill\cr
\S=\ln z-\frad{1-(1-z)\e^z}{z^2\e^z}\,\ln(1-(1-z)\e^z),\quad\hfill&
M=1-2\,\frad{1-(1-z)\e^z}{z^2\e^z}.
}
\label{sigun}
\eeq

\subsubsection*{Uncorrelated initial state}

We now investigate the distribution of the final magnetization $M_N$
for a finite chain of~$N$ spins,
with an initial state of the form~(\ref{eps}).
This magnetization is given by
\beq
NM_N=2\nu+NM_N(0)=2\nu+\sum_{n=1}^N\s_n(0),
\label{mm0}
\eeq
where $M_N(0)$ is the initial magnetization
and $\nu$ is the number of spin flips during the history of the system.
The final magnetization $M_N$ is therefore random in two respects,
as it depends both on the initial spin configuration
and on the numbers of spin flips during the history of each cluster.

We again introduce the characteristic function
\beq
\psi_N(\lam)=\mean{\exp(\lam NM_N)}
=\bigmean{\exp\left(2\lam\nu+\lam\sum_{n=1}^N\s_n(0)\right)},
\label{psidef}
\eeq
as well as the generating series
\be
\Psi(x,\lam)=\sum_{N\ge1}\psi_N(\lam)x^N.
\ee

The brackets in the right-hand side of~(\ref{psidef}) involve:

\noindent (i)
averaging over stochastic histories with a fixed initial configuration,

\noindent (ii)
averaging over the distribution~(\ref{eps}) of initial configurations.

The outcome after (i) is that the right-hand side of~(\ref{psidef})
is a multiplicative cluster quantity of the type investigated in Appendix~A,
where the contributions of clusters of up and down spins read
\beq
f_L=\e^{\lam L},\qquad g_L=\e^{-\lam L}\phi_L(2\lam).
\label{flgl}
\eeq
Step (ii) can now be performed.
The generating series corresponding to~(\ref{flgl}) are
\beq
f(x)=\frac{x\elam}{1-x\elam},\qquad g(x)=\Phi(x\emam,2\lam)-1.
\label{fg}
\eeq
Using~(\ref{phi}) and~(\ref{q}), we obtain
\beq
\Psi(x,\lam)=\frac{x\elam\left(\exp(\p x\elam)+(1-\p)(\e^{2\lam}-1)\right)}
{(1-x\elam)\exp(\p x\elam)+(1-(1-\p)x\elam)(\e^{2\lam}-1)}.
\label{psi}
\eeq

This expression provides the distribution of the final magnetization,
for any system size $N$ and any value of the parameter $\p$
characterizing the initial state.

By expanding~(\ref{psi}) around $\lam=0$,
we obtain generating series for the moments of $NM_N$.
The first of these series yields an expression similar to~(\ref{numean})
for $N\mean{M_N}$, with a leading term, linear in $N$,
in agreement with the expression~(\ref{mresinf}) of $M(\infty)$,
a constant boundary term, and an oscillating, factorially decaying correction.
Similarly, we find
\be
N\var M\approx 4\p\e^{-\p}((2\p^2-\p+2)\e^{-\p}-1).
\ee
For a random initial configuration $(p=1/2)$,
we have therefore $N\var M\approx 4\e^{-1}-2\e^{-1/2}=0.258456$.

Table~1 provides a comparison between exact dynamical results
for a random initial state ($p=1/2$)
and prediction of the full a priori ensemble,
concerning the main characteristics
(mean value and scaled variance) of the final energy of the three models
considered in this work.

\begin{table}[htb]
\begin{center}
\begin{tabular}{|c|c|c|c|c|}
\hline
Model$^{\vd^{\vd^\vm}}_{\vd_{\vd_\vm}}$
&$\matrix{E\cr\hbox{dynamical}}$&$\matrix{E\st\cr\hbox{a priori}}$&
$\matrix{N\var E\cr\hbox{dynamical}}$&$\matrix{N\var E\cr\hbox{a priori}}$\\
\hline
CIC		&$-0.393469$	&$-0.447214$	&$0.258456$	&$0.357771$\\
Glauber		&$-0.632121$	&$-0.447214$	&$0.406006$	&$0.357771$\\
Kawasaki	&$-0.274087$	&$-0.236840$	&$0.459839$	&$0.527638$\\
\hline
\end{tabular}
\caption{\small
Mean value and limit scaled variance of the final energy:
comparison of the exact dynamical results for a random initial condition
($p=1/2)$ with the prediction of the full a priori ensemble.}
\end{center}
\end{table}

The tails of the distribution of $M_N$
are again described by an exponential estimate of the form~(\ref{sig}):
\be
P(M_N)\sim\exp(-N\,\S_\p(M_N)),
\ee
where the large-deviation function $\S_\p(M)$ reads, in parametric form:
\beqa
&&{\hskip -5mm}
\S_\p=\ln z-\frac{(1-(1-\p)z)(1-(1-\p)z-(1-z)\e^{\p z})}
{\p z^2(2-\p-(1-\p)z)\e^{\p z}}
\ln\frac{1-(1-\p)z-(1-z)\e^{\p z}}{1-(1-\p)z},\nonumber\\
&&{\hskip -5.5mm}
M=1-2\,\frac{(1-(1-\p)z)(1-(1-\p)z-(1-z)\e^{\p z})}
{\p z^2(2-\p-(1-\p)z)\e^{\p z}}.
\label{sigeps}
\eeqa
This function has finite limits
\be
\S_\p(0)=-\frac{1}{2}\ln\frac{\p(2-\p)}{2},\qquad\S_\p(1)=-\ln(1-\p),
\ee
at the minimum magnetization $M=0$, corresponding to $z=0$,
and at the ground-state magnetization $M=1$, corresponding to $z=1/(1-\p)$.
Furthermore, the result~(\ref{sigun}) is recovered
by setting $p=1$ in~(\ref{sigeps}), as it should be.

\begin{figure}[htb]
\begin{center}
\includegraphics[angle=90,width=.7\linewidth]{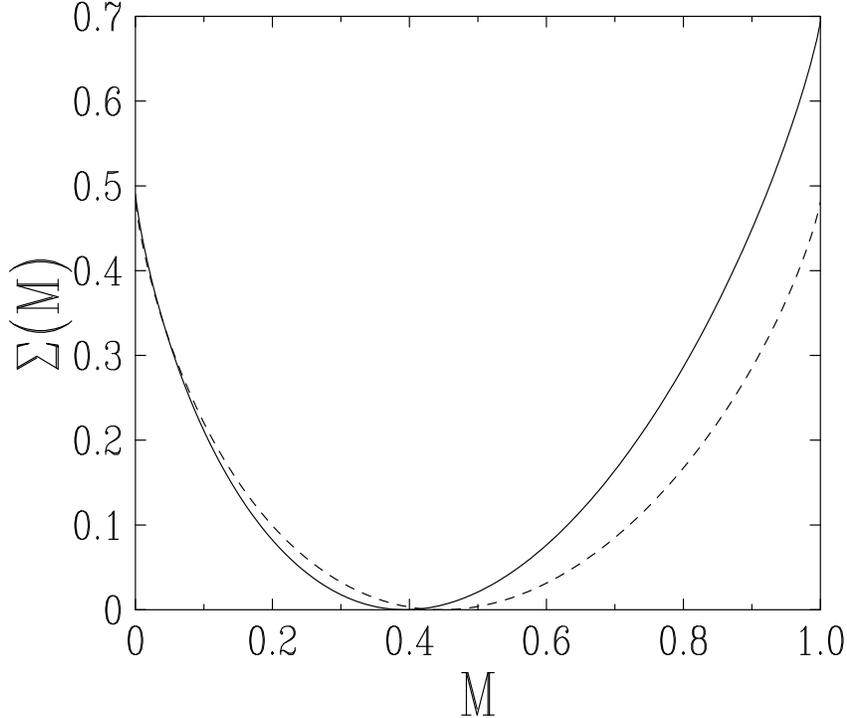}
\caption{\small
Full line: plot of the dynamical entropy of the CIC,
given by $\S_{1/2}(M)$~(\ref{sigeps}), against magnetization $M$.
Dashed line: prediction~(\ref{sm}),~(\ref{ssig}) of the a priori approach.}
\label{f4}
\end{center}
\end{figure}

Figure~\ref{f4} shows a plot of the {\it dynamical entropy},
defined as being the large-deviation function $\S_\p(M)$
of~(\ref{sigeps}) for a random initial configuration, i.e., $\p=1/2$,
against the magnetization~$M$ of the final state.
The endpoint values read
$\S_{1/2}(0)=\ln(8/3)/2=0.490415$ and $\S_{1/2}(1)=\ln 2=0.693147$.
The prediction~(\ref{ssig}) of the a priori approach is plotted for comparison.
The functions $\S_{1/2}(M)$ and $\S\ap(M)$
respectively vanish for $M(\infty)$~(\ref{minfran}) and $M\st$~(\ref{m}).
These numbers are listed in Table~1,
together with the corresponding limit scaled variances.

\section{Constrained Glauber dynamics}
\subsection{Definition of the model}

We now consider a ferromagnetic Ising chain with Glauber dynamics
(non-conserved order parameter) in the presence of kinetic constraints.
The Hamiltonian of the chain, with unit exchange constant, reads
\beq
\H=-\sum_n\s_n\s_{n+1}=-\sum_n s_n,
\label{hamf}
\eeq
where we have introduced the energy (bond) variables $s_n=\s_n\s_{n+1}$.

We consider single spin-flip (Glauber) dynamics,
assuming that the flipping rate only depends on the energy difference
between the configurations after and before the proposed move, i.e.,
\be
W(\s_n\to-\s_n)=\W_{\delta\H},
\ee
with
\be
\delta\H=2(\s_{n-1}+\s_{n+1})\s_n=2(s_{n-1}+s_n)\in\{-4,0,4\}.
\ee
The requirement that the dynamics obeys detailed balance
with respect to the Hamiltonian~(\ref{hamf}) at temperature $T=1/\beta$
yields a single condition:
\be
\frac{\W_4}{\W_{-4}}=\e^{-4\beta}.
\ee
Choosing time units such that $\W_{-4}=1$, we have $\W_4=\e^{-4\beta}$.
We restrict ourselves to zero-temperature dynamics, so that $\W_4=0$.
The rate $\W_0$, corresponding to diffusive rearrangements at constant energy,
remains a free parameter.
The zero-temperature limits of the
Metropolis and heat-bath rules respectively correspond
to $\W_0=1$ and $\W_0=1/2$.
Here we choose
\beq
\W_0=0,
\label{w00}
\eeq
so that only spin flips which lower the energy are allowed.
The condition~(\ref{w00}) defines the constrained Glauber dynamics
already considered in~\cite{ldean,prbr}.
The possible spin moves are flips of isolated spins:
\beq
-+-\to---,\qquad+-+\to+++.
\label{gls}
\eeq
Each move suppresses two consecutive unsatisfied bonds:
$s_{n-1}=s_n=-1\to s_{n-1}=s_n=+1$.
The system eventually reaches a blocked state, where there is no isolated spin.
Equivalently, each unsatisfied bond (or domain wall) is isolated.
Our aim is again to provide a statistical description
of the blocked states reached in this way.

We recast the problem in terms of deposition,
where empty sites represent unsatisfied bonds,
while occupied sites represent satisfied bonds:
\beq
\left\{\matrix{
s_n=\s_n\s_{n+1}=-1\lra\bl,\hfill\cr
s_n=\s_n\s_{n+1}=+1\lra\no.\hfill
}\right.
\label{corrbond}
\eeq
The moves~(\ref{gls}) read
\be
\bl\bl\to\no\no,
\ee
so that the dynamics is equivalent to the RSA of dimers,
considered long ago~\cite{flory,core}.

The blocked states
are the spin configurations where unsatisfied bonds are isolated.
These blocked configurations are therefore formally
equivalent to those of the CIC of section~2,
up to the replacement of the spins $\s_n$ by the energy variables $s_n$.
The Hamiltonians~(\ref{ham}) and~(\ref{hamf})
are also equivalent, up to the replacement $\s_n\to s_n$.
As a consequence, the entropy $S\ap(E)$ of the a priori ensemble
at fixed energy $E$ is still given by~(\ref{sm}),
up to the replacement of $M$ by $-E$.

\subsection{Dynamics of cluster densities and energy}

We again consider an initial state similar to~(\ref{eps}),
with $\s_0(0)=\pm1$ at random,
while each energy variable is drawn from the binary distribution
\beq
\left\{\matrix{
s_n(0)=-1\hfill&(\bl)\hfill&\hbox{with prob.}\;\p,\hfill\cr
s_n(0)=+1\hfill&(\no)\hfill&\hbox{with prob.}\;1-\p.\hfill
}\right.
\label{epsf}
\eeq
The parameter $\p$ is related to the initial energy $E(0)=-1+2\p$,
and~(\ref{beta0}) still holds.

The dynamics of the cluster densities and energy
can be investigated by the method of section~2.3.
The densities $p_\ell(t)$ of clusters of exactly~$\ell$
consecutive unsatisfied bonds (empty sites) obey linear equations
similar to~(\ref{prat}):
\be
\frac{\d p_\ell(t)}{\d t}=-(\ell-1)p_\ell(t)+2\sum_{k\ge\ell+2}p_k(t)
\ee
for $\ell\ge1$, with $p_\ell(0)=(1-\p)^2\p^\ell$, and the energy reads
\beq
E(t)=-1+2\sum_{\ell\ge1}\ell\,p_\ell(t).
\label{edefp}
\eeq
The Ansatz~(\ref{pan}) again holds, yielding the solution
\be
p_\ell(t)=(1-\p\e^{-t})^2\exp(2\p(\e^{-t}-1))\,\p^\ell\e^{-(\ell-1)t}
\ee
and
\be
E(t)=-1+2\p\exp(2\p(\e^{-t}-1)).
\ee
Again, only inactive clusters of length $\ell=1$ survive in the blocked states:
\be
p_1(\infty)=\p\e^{-2\p},
\ee
so that
\beq
E(\infty)=-1+2\p\e^{-2\p}.
\label{efinf}
\eeq
This result~\cite{ldean,prbr} was shown in Figure~\ref{f1}.

For an initial state close to the ferromagnetic ground-state
($E(0)\to-1$, i.e., $\p\to0$), the behavior $E(\infty)\approx E(0)-4\p^2$
is easily explained in terms of clusters of two unsatisfied bonds.
The energy of blocked states then increases monotonically against $\p$,
up to the maximum value
\be
E(\infty)_{\p=1/2}=-1+\e^{-1}=-0.632121,
\ee
corresponding to a random initial configuration ($\p=1/2$, i.e., $E(0)=0$),
and then decreases monotonically against $\p$, down to the value
\be
E(\infty)_{\p=1}=-1+2\e^{-2}=-0.729329,
\ee
corresponding to the antiferromagnetically ordered initial state ($\p=1$).

\subsection{Spin and energy correlations}

In the present context, it is natural to consider the spin (site)
and energy (bond) correlation functions
\be
C_n(t)=\mean{\s_0\s_n}_t,\qquad
\G_n(t)=\mean{s_0s_n}_t=\mean{\s_0\s_1\s_n\s_{n+1}}_t.
\ee

The energy correlation function $\G_n(t)$ can be evaluated analytically,
using the method of section~2.4.
We introduce variables $\t_n=(1-s_n)/2$,
and consider the one-cluster function $c_n(t)$
and the two-cluster function $d_{m,k,n}(t)$,
defined in~(\ref{cn}) and~(\ref{dn}).
These functions obey rate equations similar to~(\ref{crat}) and~(\ref{drat}):
\bea
&&{\hskip .63cm}\frac{\d c_n(t)}{\d t}=-(n-1)c_n(t)-2c_{n+1}(t),\nonumber\\
&&\frac{\d d_{m,k,n}(t)}{\d t}=-(m+n-2)d_{m,k,n}(t)\nonumber\\
&&{\hskip 2.25cm}-d_{m+1,k,n}(t)-d_{m+1,k-1,n}(t)
-d_{m,k-1,n+1}(t)-d_{m,k,n+1}(t).
\eea

\begin{figure}[htb]
\begin{center}
\includegraphics[angle=90,width=.7\linewidth]{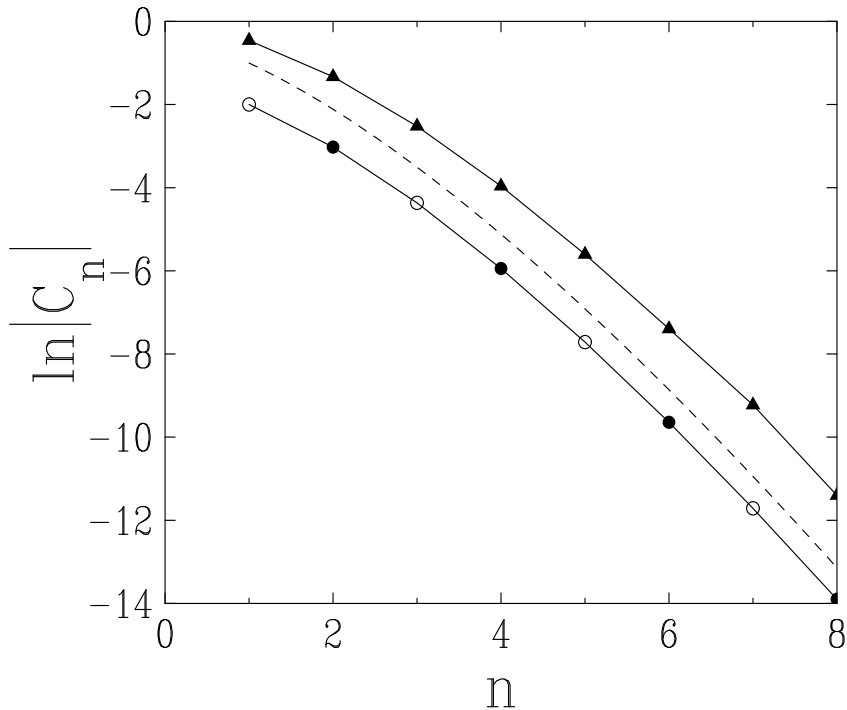}
\caption{\small
Spin and energy correlation function in the blocked states
of the ferromagnetic chain with constrained Glauber dynamics.
Full (open) symbols show positive (negative) correlations.
Circles and full line: logarithm of $(-1)^n$ times
the connected energy correlation $\G^\con_n(\infty)$~(\ref{cfinf})
for $\p=1/2$, against $n$.
Triangles and full line: logarithm of the full
spin correlation function $C_n(\infty)$, against $n$,
measured in a numerical simulation.
Dashed line: logarithm of asymptotic behavior $1/(n+1)!$,
up to a multiplicative constant, meant as a guide to the eye.}
\label{f5}
\end{center}
\end{figure}

After some algebra we are left with the following expression
for the connected energy correlation function $\G^\con_n(\infty)$:
\beq
\G^\con_n(\infty)=\G_n(\infty)-E(\infty)^2
=2\p\e^{-2\p}\left((1-2\p)\frac{(-2\p)^n}{n!}
-2\p\sum_{m\ge n+1}\frac{(-2\p)^m}{m!}\right),
\label{cfinf}
\eeq
which closely resembles~(\ref{cresinf}).
For $\p\ne1/2$, the first term is leading,
hence $\G^\con_n(\infty)\sim(-2\p)^n/n!$.
For $\p=1/2$, $\G^\con_n(\infty)\sim(-1)^n/(n+1)!$.
Figure~\ref{f5} shows a logarithmic plot of both correlation functions,
against $n$, for a random initial configuration $(\p=1/2)$.
The circles show $(-1)^n\,\G^\con_n(\infty)$,
as given by the analytical result~(\ref{cfinf}).
The triangles show the full spin correlation function $C_n(\infty)$,
measured in a numerical simulation.
For each sample, starting in a random initial configuration,
the constrained dynamics is run until a blocked state is reached.
The correlation function $C_n(\infty)$ is found to be positive
and to decay monotonically to zero as a function of the distance $n$.
The data shown correspond to a total of $10^{10}$ blocked spins.
Both correlations are observed to fall off as $1/(n+1)!$.

\subsection{Distribution of final energy and dynamical entropy}

We now investigate the distribution of the number of spin flips
and of the final energy, using the method of section~2.5.

We consider first the case of a single cluster of size $\ell\ge2$,
whose initial configuration is antiferromagnetically ordered,
i.e., made of unsatisfied bonds.
Let $\nu_\ell$ be the number of spin flips during the history of this cluster.
Equation~(\ref{nucut}) is replaced by
\be
\nu_\ell=2+\nu_{\ell_1}+\nu_{\ell_2},
\ee
where $\ell_1=n-1$ and $\ell_2=\ell-n-1$,
and the breaking point $n$ is uniform in the range $1\le n\le\ell-1$.
Hence equation~(\ref{ricc}) for the generating series $\Phi(x,\lam)$
is replaced by
\beq
x\frac{\d\Phi(x,\lam)}{\d x}=\left(x\elam\Phi(x,\lam)\right)^2
+\Phi(x,\lam)-1.
\label{riccf}
\eeq
This is a Riccati equation, which can be solved by linearization~\cite{ode}.
Setting
\be
\frac{1}{\Phi(x,\lam)}=1-\frac{x}{u(x,\lam)}\,\frac{\d u(x,\lam)}{\d x}
\ee
yields
\be
\frac{\d^2u(x,\lam)}{\d x^2}=\e^{2\lam}u(x,\lam),
\ee
so that a basis of solutions reads $\exp(\pm x\elam)$.
We thus obtain the closed-form expression
\be
\Phi(x,\lam)=\frac{(\elam+1)\exp(2x\elam)+\elam-1}
{(\elam+1)(1-x\elam)\exp(2x\elam)+(\elam-1)(1+x\elam)}.
\ee

We now consider the final total energy $NE_N$
of a finite chain of $N$ spins, with the initial state~(\ref{epsf}).
Equation~(\ref{mm0}) is replaced by
\be
NE_N=NE_N(0)-2\nu,
\ee
where $E_N(0)$ is the initial energy and $\nu$ is the number of spin flips.
Equations~(\ref{fg}) and~(\ref{psi}) are replaced by
\be
f(x)=\frac{x\emam}{1-x\emam},\qquad g(x)=\Phi(x\elam,-2\lam)-1
\ee
and
\be
\Psi(x,\lam)=
\frac{x\left((\e^{2\lam}+1)\exp(2\p x\emam)+(2\p-1)(\e^{2\lam}-1)\right)}
{(\e^{2\lam}+1)(\elam-x)\exp(2\p x\emam)+(-\elam+(1-2\p)x)(\e^{2\lam}-1)}.
\ee

This last expression contains the distribution of the final energy
of a system of size $N$,
as a function of the parameter $\p$ characterizing the initial state.
In particular, the mean energy is found to agree
with the expression~(\ref{efinf}) of $E(\infty)$,
while its scaled variance reads
\be
N\var E\approx 4\p(4\p^2-\p+1)\e^{-4\p}.
\ee
For a random initial configuration $(p=1/2)$,
we have therefore $N\var E\approx 3\e^{-2}=0.406006$.

\begin{figure}[htb]
\begin{center}
\includegraphics[angle=90,width=.7\linewidth]{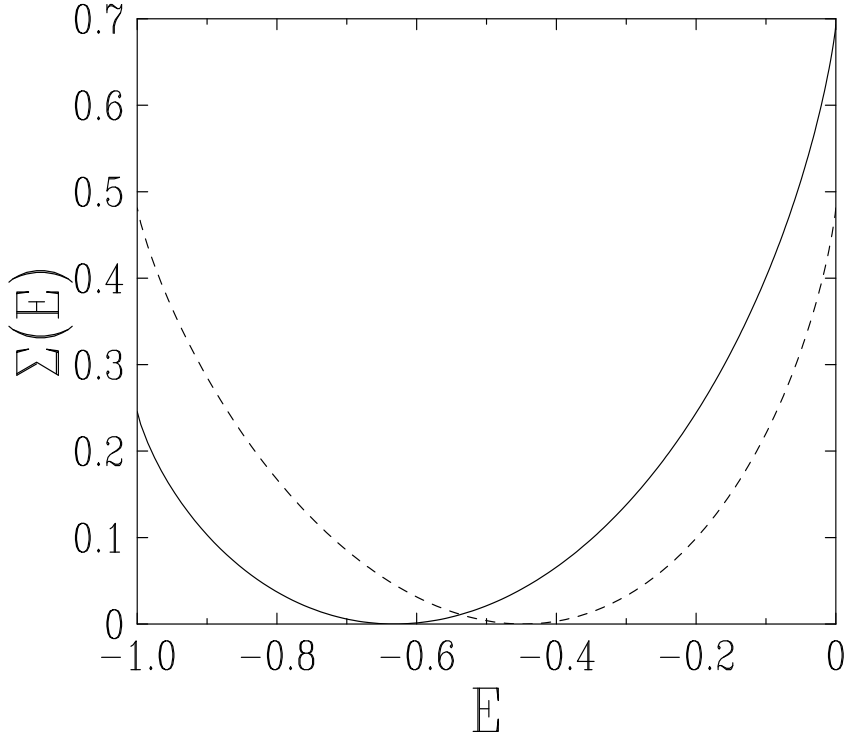}
\caption{\small
Full line: plot of the dynamical entropy
of the ferromagnetic chain with constrained Glauber dynamics,
given by $\S_{1/2}(E)$~(\ref{sigf}), against energy $E$.
Dashed line: prediction~(\ref{sm}),~(\ref{ssig}) of the a priori approach.}
\label{f6}
\end{center}
\end{figure}

The tails of the distribution of $E_N$
are again given by an estimate similar to~(\ref{sig}):
\be
P(E_N)\sim\exp(-N\,\S_\p(E_N)),
\ee
where the large-deviation function $\S_\p(E)$ reads, in parametric form:
\beqa
&&{\hskip -5mm}
\S_\p=\ln z+\frac{(1+(2\p-1)z)^2-(z-1)^2\e^{4\p z}}
{4\p z^2\e^{2\p z}(2(1-\p)+(2\p-1)z)}
\ln\frac{1+(2\p-1)z+(1-z)\e^{2\p z}}{1+(2\p-1)z-(1-z)\e^{2\p z}},\nonumber\\
&&{\hskip -3.8mm}
E=-1+\frac{(1+(2\p-1)z)^2-(z-1)^2\e^{4\p z}}
{2\p z^2\e^{2\p z}(2(1-\p)+(2\p-1)z)}.
\label{sigf}
\eeqa
This function has finite limits
\be
\S_\p(-1)=\ln z_c(\p),\qquad\S_\p(0)=-\frac{1}{2}\ln(\p(1-\p)),
\ee
at the ground-state energy $E=-1$, corresponding to $z=z_c(\p)$, with
\beq
1+(2\p-1)z_c+(1-z_c)\e^{2\p z_c}=0,
\label{zcp}
\eeq
and at the maximum energy $E=0$, corresponding to $z\to0$.
Figure~\ref{f6} shows a plot of the dynamical entropy $\S_{1/2}(E)$,
as given by~(\ref{sigf}), against the energy $E$ of the final state.
The prediction of the a priori approach is plotted for comparison.
The endpoint values of the dynamical entropy
are $\S_{1/2}(-1)=\ln z_c(1/2)=0.245660$ and $\S_{1/2}(0)=\ln 2=0.693147$.

\section{Constrained Kawasaki dynamics}
\subsection{Definition of the model}

We finally investigate a ferromagnetic Ising chain with conserved dynamics
at zero temperature, in the presence of kinetic constraints.

Consider the ferromagnetic chain with Hamiltonian~(\ref{hamf}),
but now with Kawasaki dynamics,
where only pairs of opposite spins ($s_n=\s_n\s_{n+1}=-1$) may be flipped,
so that the magnetization is locally conserved.
The flipping rates are again assumed to depend only on
the energy difference involved, which now reads
\be
\delta\H=2(\s_{n-1}\s_n+\s_{n+1}\s_{n+2})
=2(s_{n-1}+s_{n+1})\in\{-4,0,4\}.
\ee
We make the choice~(\ref{w00}), defining thus a model with
constrained Kawasaki dynamics, already considered in~\cite{priv,lint,kra}.
The possible spin moves are
\beq
-+-+{}\to--++,\qquad+-+-{}\to++--.
\label{kas}
\eeq
Each move suppresses two unsatisfied bonds:
$s_{n-1}=s_{n+1}=-1\to s_{n-1}=s_{n+1}=+1$.
The system eventually reaches a blocked state,
where the spin patterns $+-+-{}$ and $-+-+{}$ are absent.
Equivalently, there are at most two consecutive unsatisfied bonds.

Considering unsatisfied bonds as empty and satisfied bonds as occupied,
as in~(\ref{corrbond}), the moves~(\ref{kas}) read
\be
\bl\bl\bl\to\no\bl\no,
\ee
so that the dynamics is equivalent to the random deposition of hollow trimers,
a case of RSA that seems not to have been studied so far.

\subsection{A priori statistics}

We consider the restricted a priori ensemble of blocked configurations
with energy~$E$.
The entropy $S\ap(E)$ of this ensemble can again
be evaluated by means of the transfer-matrix formalism.
The partition functions of a finite chain of $N$ spins,
now labeled by the prescribed values
of its last two bonds ($\bl$ or $\no$), obey the recursion
\be
\pmatrix
{Z_{N+1}^{\no\no}\cr Z_{N+1}^{\no\bl}\cr Z_{N+1}^{\bl\no}\cr Z_{N+1}^{\bl\bl}}
=\T\pmatrix{Z_N^{\no\no}\cr Z_N^{\no\bl}\cr Z_N^{\bl\no}\cr Z_N^{\bl\bl}},
\ee
where the $4\times4$ transfer matrix
\be
\T=\pmatrix{\e^\beta&0&\e^\beta&0\cr\e^{-\beta}&0&\e^{-\beta}&0\cr
0&\e^\beta&0&\e^\beta\cr0&\e^{-\beta}&0&0}
\ee
has a reducible characteristic polynomial, $\lam\,P_3(\lam)$,
with $P_3(\lam)=\lam^3-\e^\beta\lam^2-\lam-\e^{-\beta}$.
Thermodynamic quantities are still given by~(\ref{thertrans})
in terms of the largest root $\lam_+$ of $P_3$, i.e., in parametric form:
\be
z=\e^\beta\lam_+,\qquad\lam_+^2=\frac{z^2+z+1}{z},
\qquad\e^{2\beta}=\frac{z^3}{z^2+z+1},
\ee
hence
\beq
E=\frac{1-z^2}{z^2+2z+3},\qquad
S\ap=\frac{(z^2+z+1)\ln(z^2+z+1)-z(2z+1)\ln z}{z^2+2z+3}.
\label{eke}
\eeq

The entropy of the full a priori ensemble of blocked states,
irrespective of their energy,
is equal to the maximum value of the entropy $S\ap(E)$,
corresponding to $\beta=0$, where we have $z^3-z^2-z-1=0$,
hence $z=z_0=1.839287$ and
\be
S\aps=\ln z_0=0.609378,\qquad E\st=\frac{1-z_0^2}{z_0^2+2z_0+3}=-0.236840.
\ee
The difference $\S\ap(E)=S\aps-S\ap(E)$, introduced in~(\ref{ssig}),
to be plotted in Figure~\ref{f8}, vanishes quadratically as
\be
\S\ap(E)\approx c\left(E-E\st\right)^2,
\qquad c=\frac{(z_0^2+2z_0+3)^2}{8z_0^4(z_0^2+4z_0+1)}=0.947620,
\ee
so that $N\var E\approx 1/(2c)=0.527638$.

\subsection{Dynamics of cluster densities and energy}

We again consider an initial state of the form~(\ref{epsf}).
The densities $p_\ell(t)$ obey linear equations similar to~(\ref{prat}):
\be
\frac{\d p_\ell(t)}{\d t}=-(\ell-2)p_\ell(t)+2\sum_{k\ge\ell+3}p_k(t)
\ee
for $\ell\ge2$, with $p_\ell(0)=(1-\p)^2\p^\ell$.
The Ansatz~(\ref{pan}) yields the solution
\beq
p_\ell(t)=(1-\p\e^{-t})^2\,
\exp\Big(2\p(\e^{-t}-1)+\p^2(\e^{-2t}-1)\Big)\,\p^\ell\e^{-(\ell-2)t}.
\label{pell}
\eeq
The dynamical equation for $\ell=1$,
\be
\frac{\d p_1(t)}{\d t}=p_3(t)+\sum_{k\ge4}kp_k(t),
\ee
is special, because any move generates an isolated unsatisfied bond.
Using~(\ref{pell}), we get
\beq
p_1(t)=\p+\p^2(\p\e^{-t}-2)\exp\Big(2\p(\e^{-t}-1)+\p^2(\e^{-2t}-1)\Big)
-2\p^2\e^{-(1+\p)^2}\int^{1+\p}_{1+\p\e^{-t}}\e^{y^2}\,\d y.
\label{pun}
\eeq
Equation~(\ref{edefp}) then yields
\be
E(t)=-1+2\p-4\p^2\e^{-(1+\p)^2}
\int^{1+\p}_{1+\p\e^{-t}}\e^{y^2}\,\d y.
\ee
Only inactive clusters of one or two unsatisfied bonds
survive in the final states:
\be
p_1(\infty)=\p-2\p^2\e^{-2\p-\p^2}
-2\p^2\e^{-(1+\p)^2}\int^{1+\p}_1\e^{y^2}\,\d y,\qquad
p_2(\infty)=\p^2\e^{-2\p-\p^2},
\ee
so that
\beq
E(\infty)=-1+2\p-4\p^2\e^{-(1+\p)^2}\int^{1+\p}_1\e^{y^2}\,\d y.
\label{ekinf}
\eeq
This result was shown in Figure~\ref{f1}.

For an initial state close to the ferromagnetic ground-state
($E(0)\to-1$, i.e., $\p\to0$), the behavior $E(\infty)\approx E(0)-4\p^3$
is easily explained in terms of clusters of three unsatisfied bonds.
The energy of blocked states then increases monotonically against $\p$,
to the maximum value
\be
E(\infty)_{\p=1}=1-4\e^{-4}\int^2_1\e^{y^2}\,\d y=-0.098204,
\ee
corresponding to the antiferromagnetically ordered initial state ($\p=1$).
For a random initial configuration ($\p=1/2$), we have
\be
E(\infty)_{\p=1/2}=-\e^{-9/4}\int^{3/2}_1\e^{y^2}\,\d y=-0.274087.
\ee
The last two results are already in~\cite{priv,kra}.

\subsection{Spin and energy correlations}

The energy correlation function $\G_n(\infty)$
can be evaluated analytically, similarly to sections~2.4 and~3.3.
In the present situation final results are, however, less explicit.

We consider the one-cluster function $c_n(t)$
and the two-cluster function $d_{m,k,n}(t)$, defined in~(\ref{cn}),~(\ref{dn}).
The rate equations obeyed by these functions,
and the way to solve them, are similar to~(\ref{crat}),~(\ref{drat}).

The one-cluster function obeys
\bea
&&\frac{\d c_n(t)}{\d t}=-(n-2)c_n(t)-2c_{n+1}(t)-2c_{n+2}(t)
\qquad(n\ge2),\nonumber\\
&&\frac{\d c_1(t)}{\d t}=-2c_3(t),
\eea
hence
\bea
&&c_n(t)=\exp\Big(2\p(\e^{-t}-1)+\p^2(\e^{-2t}-1)\Big)\,p^n\e^{-(n-2)t}
\qquad(n\ge2),\nonumber\\
&&c_1(t)=\p-2\p^2\e^{-(1+\p)^2}\int^{1+\p}_{1+\p\e^{-t}}\e^{y^2}\,\d y,
\nonumber
\eea
so that
\beq
c_1(\infty)=\p-2\p^2\e^{-(1+\p)^2}\int^{1+\p}_1\e^{y^2}\,\d y,
\qquad c_2(\infty)=\p^2\e^{-2\p-\p^2},
\label{kcinf}
\eeq
in agreement with~(\ref{pc}),~(\ref{pell}), and~(\ref{pun}).

The two-cluster function obeys
\beqa
&&\frac{\d d_{m,k,n}(t)}{\d t}
=-(m-2)d_{\act{m},k,n}(t)-(n-2)d_{m,k,\act{n}}(t)\nonumber\\
&&{\hskip 2.25cm}-d_{\act{m+1},k,n}(t)-d_{\act{m+1},k-1,n}(t)
-d_{m,k-1,\act{n+1}}(t)-d_{m,k,\act{n+1}}(t)\label{edk}\\
&&{\hskip 2.25cm}-d_{\act{m+2},k,n}(t)-d_{\act{m+2},k-2,n}(t)
-d_{m,k-2,\act{n+2}}(t)-d_{m,k,\act{n+2}}(t).\nonumber
\eeqa
Besides the conventions of section~2.4, only some of the terms
are present in the right-hand side for either $m$ or $n=1$ or 2,
namely those for which the underlined index is greater than 2.
Furthermore, for $k=1$ the sum $d_{\act{m+2},k-2,n}(t)+d_{m,k-2,\act{n+}2}(t)$
is replaced by $c_{m+n+1}(t)$.
As in previous cases, we look for a solution to~(\ref{edk}) of the form
\bea
&&d_{m,k,n}(t)=A_k(t)z(t)^{m+n}\qquad(m,n\ge2),\nonumber\\
&&{\hskip .03cm}d_{m,k,1}(t)=B_k(t)z(t)^m{\hskip 1.23cm}(m\ge2),\nonumber\\
&&{\hskip .14cm}d_{1,k,1}(t)=D_k(t),\nonumber
\eea
where $z(t)$ has been introduced in~(\ref{z}).
The procedure then consists in introducing
generating series $\A(x,t)$, $\B(x,t)$, $\D(x,t)$, similar to~(\ref{genera}),
writing differential equations obeyed by these functions,
and solving the latter equations.
This requires some lengthy and tedious algebra.

The energy correlation function $\G_n$ in the blocked states
is still given by~(\ref{cord})
in terms of $c_1(\infty)$ and $d_{1,n-1,1}(\infty)$,
so that the function of most interest is $\D(x,\infty)$,
for which we are left with the expression
\beqa
&&{\hskip -.64cm}\D(x,\infty)=\frac{\p^2}{1-x}
-\p^3(x+2)\e^{-(1+\p)^2}\int^{1+\p}_1\e^{y^2}\,\d y\nonumber\\
&&{\hskip .88cm}-\frac{2\p^3}{1-x}\sqr{2(x^2+1)}
\,\exp\left(-\frac{(x+1+\p(x^2+1))^2}{2(x^2+1)}\right)
\int_{b(0)}^{b(1)}\e^{y^2}\,\d y
\label{dfinal}\\
&&{\hskip .88cm}+\frac{\p^3}{x^2}\sqr{2(x^2+1)}
\,\exp\left(-\frac{(x+1+\p(x^2+1))^2}{2(x^2+1)}\right)
\int_0^1 R(x,u)\,\d u\int_{b(0)}^{b(u)}\e^{y^2}\,\d y,\nonumber
\eeqa
with the notations
\bea
&&R(x,u)=\frac{x^2+1}{1-x}(1-x+\p x(1-x)+2\p^2x^2)\nonumber\\
&&{\hskip 3cm}\times\exp\left(\p(x+1)(u-1)
+\frac{\p^2}{2}(x^2+1)(u^2-1)\right)\nonumber\\
&&{\hskip 1.25cm}-(1-x^2+\p x(1-2x-x^2)u)\nonumber\\
&&{\hskip 3cm}\times\exp\left(\p(1-x)(u-1)
+\frac{\p^2}{2}(1-x^2)(u^2-1)\right)\nonumber\\
&&{\hskip 1.25cm}+(x^2+1)\int^{1+\p x}_{1+\p xu}\e^{-y^2}\,\d y\nonumber\\
&&{\hskip 3cm}\times\exp\left(1+\p(x-1+(x+1)u)
+\frac{\p^2}{2}\left(x^2-1+(x^2+1)u^2\right)\right)
\eea
and
\be
b(u)=\frac{x+1+\p(x^2+1)u}{\sqr{2(x^2+1)}}.
\ee

\begin{figure}[htb]
\begin{center}
\includegraphics[angle=90,width=.7\linewidth]{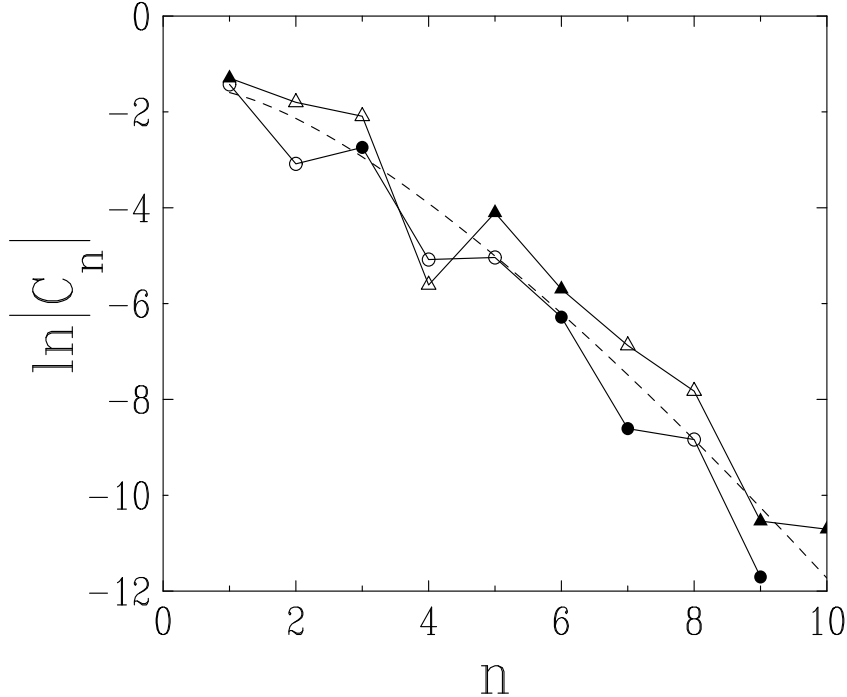}
\caption{\small
Spin (site) and energy (bond) correlation functions in the blocked states
of the ferromagnetic chain with constrained Kawasaki dynamics,
measured in a numerical simulation.
Full (open) symbols show positive (negative) correlations.
Circles and full line: logarithm of the absolute value
of the connected energy correlation $\G^\con_n(\infty)$.
Triangles and full line: logarithm of the absolute value
of the full spin correlation $C_n(\infty)$.
Dashed line: logarithm of asymptotic behavior $1/(2^n(n/2)!)$
up to a multiplicative constant, meant as a guide to the eye.}
\label{f7}
\end{center}
\end{figure}

The function $\D(x,\infty)$
has a simple pole at $x=1$, with residue $-c_1(\infty)^2$,
where $c_1(\infty)$ has been evaluated in~(\ref{kcinf}),
so that the $d_{1,k,1}(\infty)$ converge to $c_1(\infty)^2$
as the distance $k$ becomes infinitely large, as it should.
The fall-off of the difference $d_{1,k,1}(\infty)-c_1(\infty)^2$,
and that of the connected correlation function $\G^\con(\infty)$,
are related to the behavior of $\D(x,\infty)$ as $\abs{x}$ is large.
The result~(\ref{dfinal}) leads to the estimate
\be
\D(x,\infty)\sim\exp(2\p x-\p^2x^2),
\ee
with exponential accuracy.
The results summarized in Appendix~B then imply
\beq
\G^\con_n(\infty)\sim\frac{\p^n}{(n/2)!}
\cos\left(\frac{n\pi}{2}-\sqr{2n}\right).
\label{kasy}
\eeq
The result~(\ref{dfinal}) does not however lead to any useful expression
for $\G_n$, even for $n=1$.
Figure~\ref{f7} shows a logarithmic plot of the spin and energy
correlation functions against~$n$,
as measured in a numerical simulation for a random initial condition
$(\p=1/2)$.
The data shown correspond to a total of $5\cdot10^{10}$ spins.
Both the full spin correlation $C_n(\infty)$ and the connected
energy correlation $\G^\con_n(\infty)=\G_n(\infty)-E(\infty)^2$
are found to agree with the asymptotic result~(\ref{kasy}).
The absolute value of the data follows the predicted fall-off,
shown as a dashed line, while the signs roughly follow the predicted pattern
$(++--)$, up to more and more seldom mistakes.

\subsection{Distribution of final energy and dynamical entropy}

We end up with the distribution of the final energy of a finite sample.
This analysis will follow the lines of section~2.5 and~3.4,
the main difference being that~(\ref{ricck}) will have to be solved
numerically.

We consider first the case of a single cluster of size $\ell\ge2$,
whose initial configuration is only made of unsatisfied bonds.
Let $\nu_\ell$ be the number of spin flips of this cluster.
Equation~(\ref{nucut}) is replaced by
\be
\nu_\ell=2+\nu_{\ell_1}+\nu_{\ell_2},
\ee
where $\ell_1=n-1$ and $\ell_2=\ell-n-2$,
and the breaking point $n$ is uniform in the range $1\le n\le\ell-2$.
Hence equation~(\ref{ricc}) for the generating series $\Phi(x,\lam)$
is replaced by
\beq
x\frac{\d\Phi(x,\lam)}{\d x}=x^3\e^{2\lam}\Phi(x,\lam)^2+2\Phi(x,\lam)-x-2.
\label{ricck}
\eeq
In contrast with~(\ref{ricc}) and~(\ref{riccf}),
we have not been able to solve the Riccati equation~(\ref{ricck}) analytically.

We now consider the final total energy $NE_N$
of a finite chain of $N$ spins, with the initial state~(\ref{epsf}).
The tails of the distribution of $E_N$
are again given by an estimate similar to~(\ref{sig}),
\be
P(E_N)\sim\exp(-N\,\S_\p(E_N)),
\ee
where the large-deviation function $\S_\p(E)$ reads, in parametric form,
\beq
E=-\frac{1}{x_c}\frac{\d x_c}{\d\lam},\qquad
\S_\p=\ln x_c-\frac{\lam}{x_c}\frac{\d x_c}{\d\lam},
\label{sk}
\eeq
and $x_c(\lam)=\exp(-F(\lam))$ is the real positive solution of
\beq
\Phi(\p x_c\elam,-2\lam)=\frac{\elam}{(1-\p)x_c}.
\label{xck}
\eeq

\begin{figure}[htb]
\begin{center}
\includegraphics[angle=90,width=.7\linewidth]{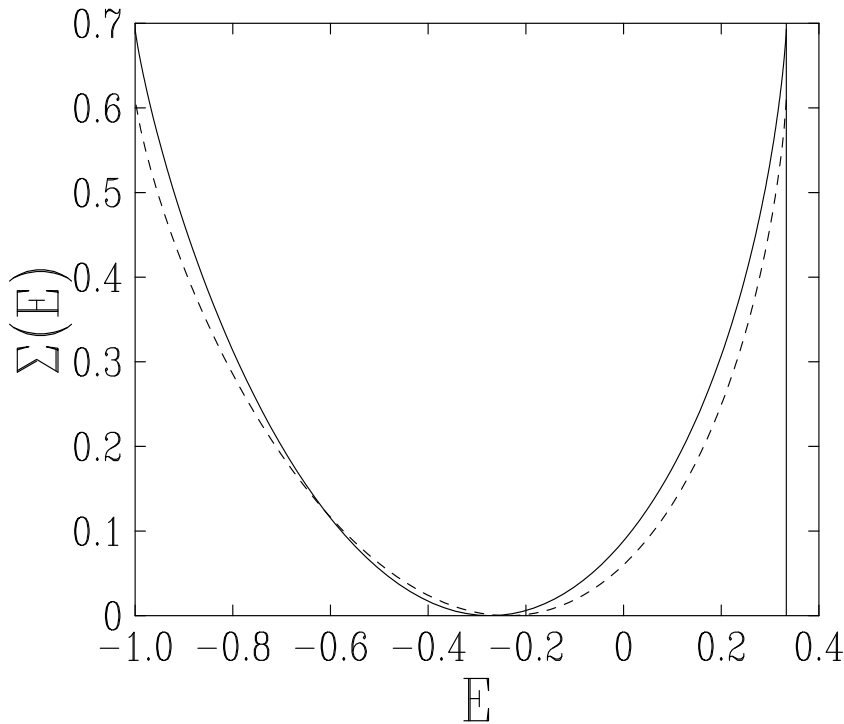}
\caption{\small
Full line: plot of the dynamical entropy
of the ferromagnetic chain with constrained Kawasaki dynamics,
against energy $E$.
Data are obtained by solving numerically
(\ref{ricck}), (\ref{sk}), and (\ref{xck}), for $\p=1/2$.
Dashed line: prediction~(\ref{eke}) of the a priori approach.}
\label{f8}
\end{center}
\end{figure}

The function $\S_\p(E)$ can be evaluated analytically in some regimes.
Skipping any detail, we mention that the limits
\be
\S_\p(-1)=-\ln(1-\p),\qquad\S_\p(1/3)=-\frac{1}{3}\ln(\p^2(1-\p)),
\ee
at the ground-state energy $E=-1$ and at the maximum energy $E=1/3$
can be determined exactly.
Moreover, the solution to~(\ref{ricck}) can be expanded as a power series
$\Phi(x,\lam)=1/(1-x)+\lam\Phi_1(x)+\lam^2\Phi_2(x)+\cdots$
We thus recover the mean energy $E(\infty)$~(\ref{ekinf}),
and obtain the following expression for the scaled energy variance:
\bea
&&N\var E\approx 2\p(1-\p)(1-4\p^2)+4(1-\p)^2B(\p)\nonumber\\
&&{\hskip 1.5cm}-4(1-\p)^2(3-4\p+4\p^3)A(\p)-2(1-\p)^3(5-7\p+4\p^3)A(\p)^2,
\eea
with
\bea
&&A(\p)=\frac{2\p^2}{(1-\p)^2}\e^{-(1+\p)^2}\int_1^{1+\p}\e^{y^2}\,\d y
=\frac{2\p-1-E(\infty)}{2(1-\p)^2},\nonumber\\
&&B(\p)=\frac{\p^2}{(1-\p)^2}\e^{-(1+\p)^2}\int_1^{1+\p}\e^{y^2}
((y-2)A(y-1)-2)^2\,\d y.
\eea
Figure~\ref{f8} shows a plot of the dynamical entropy $\S_{1/2}(E)$
obtained by solving numerically (\ref{ricck}), (\ref{sk}), (\ref{xck}).
The prediction~(\ref{eke}) of the a priori approach is shown for comparison.
For $\p=1/2$, we have $\S_{1/2}(-1)=\S_{1/2}(1/3)=\ln 2=0.693147$
and $N\var E\approx 0.459839$.

\section{Discussion}

We have presented a parallel study
of the zero-temperature dynamics of three one-dimensional
Ising models with kinetic constraints,
to which a number of previous studies have already been devoted:
paramagnetic CIC models~\cite{scic,scictau,acic,acictau}~(section~2),
ferromagnetic chain
with constrained Glauber dynamics~\cite{ldean,prbr}~(section~3),
ferromagnetic chain
with constrained Kawasaki dynamics~\cite{priv,lint,kra}~(section~4).

The common characteristic feature of these models
is that their zero-temperature dynamics is fully irreversible:
each spin can flip at most once during its whole history.
As shown in the present work, these stochastic dynamical systems can be mapped
onto models of irreversible deposition~\cite{rsa}:
CSA of monomers ($\no$) for paramagnetic CIC models,
RSA of dimers ($\no\no$) for the ferromagnetic chain
with constrained Glauber dynamics,
RSA of hollow trimers ($\no\bl\no$) for the ferromagnetic chain
with constrained Kawasaki dynamics.
This exact mapping onto RSA or CSA models
allows the analytical determination of many physical quantities.
Assuming an uncorrelated initial state
with prescribed energy (or magnetization),
we have obtained exact results for the three models,
and compared them to the predictions of the a priori approach,
testing thus the so-called Edwards hypothesis in this particular
zero-temperature framework.

We have first shown that the jamming time grows logarithmically with the
system size, up to finite fluctuations given by extreme-value statistics.
The result~(\ref{tx}),~(\ref{gum}), established for the CIC,
also holds quantitatively for the ferromagnetic chain
with constrained Glauber and Kawasaki dynamics,
with respectively $\alpha=p^2\e^{-2p}$ and $\alpha=p^3\e^{-2p-p^2}$.

There is a complete lack of ergodicity in these irreversible models.
The mean final energy
indeed bears a non-trivial dependence on the initial condition,
as depicted in Figure~\ref{f1}.
For a random initial configuration, the comparison of the exact dynamical
results for the average and variance of the final energy
with the prediction of the a priori (Edwards) approach
reveals systematic differences, which have either sign,
and an absolute value ranging up to some 20 percent (see Table~1).

The two-point spin (site) and energy (bond) correlation function
in the final states has also been evaluated,
either by analytical means or by accurate numerical simulations.
Connected correlations fall off
factorially~(see Figures~\ref{f3}, \ref{f5}, \ref{f7}),
often with an oscillating sign.
A super-exponential fall-off of correlations is indeed known
to be generically obeyed in RSA models.
Such a feature cannot be reproduced by an a priori ensemble,
where correlations generically decay exponentially,
with a finite correlation length, related to the first two eigenvalues
of the transfer matrix.

We have also determined the distribution of the energy
of the final states
beyond the Gaussian approximation.
Such a problem seems to have been tackled only once
in the RSA literature~\cite{core}.
We thus obtain large-deviation estimates
for the exponentially small tails of the distribution.
The corresponding dynamical entropy depends on
the initial energy (or magnetization).
The comparison of the result for a random initial configuration $(p=1/2)$
with the a priori approach again shows differences
at a quantitative level~(see Figures~\ref{f4}, \ref{f6}, \ref{f8}).

The results obtained so far invalidate the Edwards hypothesis
in the present situation of fully irreversible zero-temperature dynamics.
There are indeed systematic differences
between the exact dynamical expressions and
the predictions of the a priori approach,
and even qualitative discrepancies, such as the super-exponential
fall-off of correlations.

The present work also questions the existence of any simple relationship
between the landscape of metastable states
and the slow dynamics just above the dynamical phase transition.
Indeed, on the one hand, all the results on the zero-temperature dynamics
of the CIC are independent of the parameter $a$,
which interpolates between the ACIC for $a=0$ or $a=1$
and the SCIC for $a=1/2$.
On the other hand, these limiting cases are known to have different
kinds of slow dynamics in the presence of activated processes,
at low temperature.
For instance, the relaxation time to equilibrium diverges
as $\tau_\eq\sim\exp(2\beta)$ for the SCIC~\cite{scictau},
and as $\tau_\eq\sim\exp(\beta^2/(\ln 2))$ for the ACIC~\cite{acictau}.

In spite of its specificity, the present approach may also shed some new light
on other quantities and/or other situations of interest.
One example is the size distribution of ordered clusters,
which has been recently shown to be a useful tool
to test the Edwards hypothesis in spin models under tapping~\cite{bfs}.
In the present context the exact determination of
the density $f_\ell(\infty)$ of clusters of $\ell$ occupied sites
in the final states would require a lengthy calculation.
However its exponential fall off for a large cluster size
is related to the ground-state dynamical entropy as
$f_\ell(\infty)\sim\exp(-\ell\,\S_p(E=-1))$.
We thus obtain the simple estimate $f_\ell(\infty)\sim(1-p)^\ell$,
both for CIC and constrained Kawasaki dynamics,
expressing that the long ordered clusters in the final state
have to be already present in the initial state,
while the result $f_\ell(\infty)\sim z_c(p)^{-\ell}$ (see~(\ref{zcp}))
for constrained Glauber dynamics is non-trivial.

Finally, the present situation
of a quench from a disordered initial configuration
(infinite initial temperature)
can be viewed as the relaxation part of a cycle of random tapping
with infinitely high intensity.
It would also be desirable to extend at least some of our results
to the more realistic situation of a finite tapping intensity.

\subsubsection*{Acknowledgements}

Interesting discussions with Silvio Franz are gratefully acknowledged.

\newpage
\appendix
\section{Averaging a multiplicative cluster function}

Consider a finite chain of $N$ spins $\s_n=\pm1$, numbered $n=1,\dots,N$.
The chain is naturally partitioned into clusters of parallel spins.
Let $M$ be the number of clusters,
and $L_1,L_2,\dots,L_M$ be the lengths of the clusters,
with $L_1+\cdots+L_M=N$.
Assume $\s_1=+1$.
We have $\s_n=+1$ for $n=1,\dots,L_1$,
then $\s_n=-1$ for $n=L_1+1,\dots,L_1+L_2$, and so on.

A multiplicative cluster function is a quantity of the form
\be
q_N=\left\{\matrix{
f_{L_1}\,g_{L_2}\,f_{L_3}\cdots\hfill&\hbox{if}\ \s_1=+1,\hfill\cr
g_{L_1}\,f_{L_2}\,g_{L_3}\cdots\hfill&\hbox{if}\ \s_1=-1,\hfill
}
\right.
\ee
where each cluster of $L$ up spins brings a factor $f_L$,
and each cluster of $L$ down spins brings a factor $g_L$.

Averaging a quantity such as $q_N$ over a state of the form~(\ref{eps})
amounts to summing the contributions of all the partitions of $N$ into cluster
lengths $\{L_k, k=1,\dots,M\}$, with the a priori weight
\be
W(\{L_k\})=\left\{\matrix{
(1-\p)^{L_1}\,\p^{L_2}\,(1-\p)^{L_3}\cdots\hfill&\hbox{if}\ \s_1=+1,\hfill\cr
\p^{L_1}\,(1-\p)^{L_2}\,\p^{L_3}\cdots\hfill&\hbox{if}\ \s_1=-1.\hfill
}
\right.
\ee
In order to perform this summation, we introduce the generating series
\be
f(x)=\sum_{L\ge1}f_Lx^L,\qquad g(x)=\sum_{L\ge1}g_Lx^L,\qquad
Q(x)=\sum_{N\ge1}\mean{q_N}x^N.
\ee
If $\s_1=+1$, we have
\bea
&&Q(x)=\sum_{L_1,L_2,L_3,\dots}f_{L_1}((1-\p)x)^{L_1}
\;g_{L_2}(\p x)^{L_2}\;f_{L_3}((1-\p)x)^{L_3}\dots\\
&&{\hskip .88cm}=f((1-\p)x)+f((1-\p)x)g(\p x)
+f((1-\p)x)g(\p x)f((1-\p)x)+\cdots,
\eea
where the successive terms are the contributions of the partitions
of the chain into $M=1,2,3,\dots$ clusters.
Adding the contribution of the sector $\s_1=-1$,
and summing up the geometrical series, we obtain the result
\beq
Q(x)=\frac{f((1-\p)x)+g(\p x)+2f((1-\p)x)g(\p x)}{1-f((1-\p)x)g(\p x)},
\label{q}
\eeq
which interpolates between $Q(x)=f(x)$ at $\p=0$ and $Q(x)=g(x)$ at $\p=1$.

\newpage
\section{Expanding $\exp(ax-bx^2)$ as a power series}

This appendix is devoted to the power-series expansion
\be
\exp(ax-bx^2)=\sum_{n\ge0}f_n(a,b)x^n.
\ee
An identification with the generating series
of Hermite polynomials~\cite{htf}:
\be
\sum_{n\ge0}H_n(z)\frac{x^n}{n!}=\exp(2zx-x^2)
\ee
leads to
\be
f_n(a,b)=\frac{b^{n/2}}{n!}H_n\!\left(\frac{a}{2\sqrt{b}}\right).
\ee
We are mostly interested in the asymptotic behavior of the coefficients
$f_n(a,b)$ as $n$ gets large, for fixed $a$ and $b$.
The asymptotic expansion of Hermite polynomials~\cite{htf} yields
\beq
f_n(a,b)\approx\frac{b^{n/2}}{(n/2)!}\exp\!\left(\frac{a^2}{8b}\right)
\cos\!\left(\frac{n\pi}{2}-a\sqr{\frac{n}{2b}}\right).
\label{appc}
\eeq

The above estimate becomes exact for any finite $n$
in the simple situation where $a=0$, where one has straightforwardly
\be
f_{2k}(0,b)=\frac{(-b)^k}{k!},\qquad f_{2k+1}(0,b)=0.
\ee

For generic values of $a$,
the signs of the coefficients $f_n(a,b)$ are given by the cosine function
in~(\ref{appc}).
They oscillate according to the four-periodic pattern $(++--)$,
except for `mistakes' which take place more and more seldomly,
for $n\approx k^2\mu$, with
\be
\mu=\frac{\pi^2b}{2a^2}.
\ee
For $a>0$, mistakes are isolated $+$ or $-$ signs.
For $a<0$, they consist of three consecutive $+++$ or $---$ signs.

\end{document}